\documentclass[aps,prb,twocolumn,superscriptaddress,showpacs,a4paper,english,longbibliography]{revtex4-1}
\usepackage{amsfonts,epsfig}
\usepackage{amsmath}
\usepackage{epstopdf}
\usepackage{graphicx}
\usepackage{bm}
\usepackage{color}
\usepackage{amssymb}
\usepackage{times}
\usepackage{dcolumn}
\usepackage{cases}
\usepackage{txfonts}
\usepackage[english]{babel}
\usepackage{epstopdf}
\usepackage{hyperref}
\usepackage{soul}
\DeclareGraphicsExtensions{.png,.eps}
\sethlcolor{yellow}

\begin{document}

% Include your paper's title here

\title{Few-Body Systems Capture Many-Body Physics: Tensor Network Approach}

% Place the author information here.  Please hand-code the contact
% information and notecalls; do *not* use \footnote commands.  Let the
% author contact information appear immediately below the author names
% as shown.  We would also prefer that you don't change the type-size
% settings shown here.

\author{Shi-Ju Ran}
\email[Corresponding author. ]{Email: shi-ju.ran@icfo.eu}
\affiliation{ICFO-Institut de
	Ciencies Fotoniques, The Barcelona Institute of Science and
	Technology, 08860 Castelldefels (Barcelona), Spain}
\author{Angelo Piga}
\affiliation{ICFO-Institut de
	Ciencies Fotoniques, The Barcelona Institute of Science and
	Technology, 08860 Castelldefels (Barcelona), Spain}
\author{Cheng Peng}
\affiliation{Theoretical Condensed Matter Physics and Computational Materials Physics Laboratory, School of Physical Sciences, University of Chinese Academy of Sciences, Beijing 100049, China}
\author{Gang Su}
\affiliation{Theoretical Condensed Matter Physics and Computational Materials Physics Laboratory, School of Physical Sciences, University of Chinese Academy of Sciences, Beijing 100049, China}
\affiliation{Kavli Institute for Theoretical Sciences, University of Chinese Academy of Sciences, Beijing 100190, China}
\author{Maciej Lewenstein}
\affiliation{ICFO-Institut de Ciencies Fotoniques, The Barcelona Institute of Science and Technology, 08860 Castelldefels (Barcelona), Spain}
\affiliation{ICREA, Passeig Lluis Companys 23, 08010 Barcelona, Spain}

% Include the date command, but leave its argument blank.
%%%%%%%%%%%%%%%%% END OF PREAMBLE %%%%%%%%%%%%%%%%


\begin{abstract}
  Due to the presence of strong correlations, theoretical or experimental investigations of quantum many-body systems belong to the most challenging tasks in modern physics. Stimulated by tensor networks, we propose a scheme of constructing the few-body models that can be easily accessed by theoretical or experimental means, to accurately capture the ground-state properties of infinite many-body systems in higher dimensions. The general idea is to embed a small bulk of the infinite model in an ``entanglement bath'' so that the many-body effects can be faithfully mimicked. The approach we propose is efficient, simple, flexible, sign-problem-free, and it directly accesses the thermodynamic limit. The numerical results of the spin models on honeycomb and simple cubic lattices show that the ground-state properties including quantum phase transitions and the critical behaviors are accurately captured by only $\mathcal{O}(10)$ physical and bath sites. Moreover, since the few-body Hamiltonian only contains local interactions among a handful of sites, our work provides new ways of studying the many-body phenomena in the infinite strongly-correlated systems by mimicking them in the few-body experiments using cold atoms/ions, or developing novel quantum devices by utilizing the many-body features.
\end{abstract}


\maketitle

\section{Introduction}

    \textbf{Quantum many-body systems in one, two and three dimensions.}
    Investigating the ground states and low-lying states of strongly correlated quantum many-body systems is one of the most important challenges in modern physics. It lies in the centre of interest of condensed matter physics \cite{Auerbachbook,Wenbook}, atomic, molecular and optic physics \cite{Lewensteinbook}, and high energy physics \cite{Montvaybook, Wiesereview}. Fundamentally, these systems may exhibit exotic states and phenomena, such as spin liquids  \cite{QSL} and topological phases \cite{reviewTI,revTI1}. On the other hand, these systems have important applications in contemporary electronics, superconductivity \cite{PALreview}, spintronics \cite{reviewDasSarma} and, more recently, quantum information \cite{roadmap}. The future quantum technologies, i.e. quantum computers, quantum simulators and annealers, quantum metrology and sensing rely essentially exclusively on the use of strongly-correlated quantum lattice systems. However, the high complexity rising from strong correlations makes the exact solutions/diagonalizations impossible or inefficient in most cases. Numerical methods, benefiting from fast development of the computer technology, become nowadays the most frequently used tools capable of reliably studying such systems. Developments of new more efficient methods, with lower cost and higher accuracy, are therefore highly demanded.

    Technically, one-dimensional (1D) models are the simplest, although quantum fluctuations in 1D are particularly large \cite{Giamarchibook}. The 1D systems play important roles in electronics and spintronics, as they provide specific possibilities in controlling transport and reveal exotic excitations such as Majorana fermions \cite{Leo-nature, Marcus-recent}. They can be naturally viewed as the edges of two-dimensional (2D) systems, and may correspond to edge states of these 2D systems. \cite{Wenbook,reviewTI}. 2D systems are obviously more demanding numerically and experimentally, whereas from a physical point of view they can be taken as playgrounds for novel concepts and exotic states such as anyonic excitations \cite{Wenbook,FQHEbook}, frustrated antiferromagnetism \cite{Lhuillierreview,Sachdev}, spin liquids \cite{Balentsnew}, topological order and topological phase transitions, and graphene-like systems \cite{RMPgraphene}, etc.

    In principle, the three-dimensional (3D) models are even more interesting, as they are much closer to reality of our daily experience. Because of their extreme complexity, the adoption of various approximations to treat them is totally unavoidable. The three dimension is closer to the upper critical dimension, and one may expect that the mean-field theories would work well for them. A paradigmatic example is the Bose-Hubbard model, which can be nicely explored by bosonic dynamic mean-field theory (DMFT) \cite{Werner2010}. Such few, but well-controlled systems can serve as validation, calibration and benchmark for various numerical and analytical methods. Still, there are also 3D models that are extremely demanding to be understood, such as, among others, the spin ice \cite{Moessner-review} with pyrochlore lattice \cite{Gingras}, that is a highly frustrated magnet; the Fermi--Hubbard model, which is usually invoked to describe high temperature superconductivity of cuprates that consist of strongly correlated 2D planes weakly coupled in the transverse direction (\cite{PALreview}, see also \cite{Tilman} for a quantum simulation with ultracold atoms). The (3+1)D lattice gauge theories at high densities and temperature are also beyond the possibilities of the existing codes and machines. Generally, different approximate analytical methods might generate converse results, leading to unnecessary controversies in many cases. It turns out that finding reliable and efficient numerical methods to solve 3D quantum many-body problems becomes indeed imperative.

    \textbf{Tensor networks --- state-of-the-art.}
    The density matrix renormalization group (DMRG) \cite{DMRG} is widely recognized as a major breakthrough in the calculations of the ground states in 1D systems. Originally proposed as a mere numerical tool, the reformulation of the DMRG as a variational algorithm in terms of matrix product states (MPS) \cite{DMRGMPS} leads to the proposal of more general formalism, based on tensor networks (TN's) \cite{ReviewTNS,ReviewTNS1,MPSPEPS}. TNs provide a very general ansatz for the wave functions: the quantities of interest may be expressed as results of the contraction of a network of local tensors. It has rapidly evolved into a promising powerful tool to study large or even infinite size systems in two dimensions. In fact, TNs overcome most of the limitations of the standard numerical algorithms: for instance, in contrast to quantum Monte Carlo (QMC), TNs do not suffer from the notorious ``negative-sign'' problem \cite{QMCsign} and allow for an accurate access to frustrated spin systems and fermionic models away from half-filling.

    To what extent the TN is feasible depends on the amount of entanglement of the states to be simulated. The efficiency (computational memory and time) of the TN approaches is also determined by the capability of the current computers. In the standard formulations, TN works for low-entangled states such as the ground states of local and gapped Hamiltonians. For these states, an area law for the entanglement entropy holds, i.e. the entanglement entropy of a subsystem (consisting of a large, but finite block) scales with the block's boundary \cite{AreaLaw}. This fact explains the efficiency of the MPS-based algorithms in 1D. For the same reason, the MPS-based algorithms (e.g. DMRG) work well for small 2D systems, but are strongly limited when the size grows \cite{DMRG2D,ReviewDMRG2D,MPSPEPS}. By acknowledging this, many different competing approaches have been developed. Among others, a purely 2D ground-state TN ansatz, termed projected entangled pair state (PEPS), was proposed as a natural extension of MPS. PEPS fulfils the 2D area law of entanglement entropy \cite{MPSPEPS,PEPS11,PEPS12,ReviewTNS,ReviewTNS1,PEPSCritical}, while the multiscale entanglement renormalization ansatz (MERA) \cite{MERA} bears particular advantages for studying critical models.

    Within the existing TN algorithms, a lot of works were done on 2D quantum as well as 3D classical models, where the simulations consist in the contractions of 3D TN's \cite{iTEBD,MERA,PEPS11,PEPS12,SimpleUpdate,TRG,TERG,SRG,HOSRG,LoopTN,CTMRG0,CTMRG11,CTMRG12,CTMRG3D,VarPEPS,GrdPEPS,Fupdate,TN3D}. This well-known quantum-classical equivalence \cite{QCcorrespond} becomes very explicit in the TN terminology, and is utilized frequently in the TN approaches for ground-state \cite{iTEBD,PEPS11,PEPS12,SimpleUpdate} and thermodynamic \cite{TMRG,LTRG,ODTNS,NCD,FTPEPS} studies on discrete and even continuous \cite{cTN} systems. However, for 3D ground-state simulations, we are essentially facing the contractions of four-dimensional TN's \cite{TN4D,SimpleTN3D}, which are hardly treatable even with small bond dimensions. Therefore, developing efficient 3D quantum algorithms are strongly desired, in particular for infinite quantum systems.

    \textbf{``Bath-stimulated'' methods.}
    For 3D quantum models, many interesting issues remain to be explored or even unsolved to a large extent \cite{QSL,SpinGlass3D,SpinIceRev,Frustration1,Frustration2}. They have been the subject of intensive studies in recent years and many numerical methods were developed to handle them. Several approaches were proposed beyond the standard mean-field and renormalization group methods, such as the linked cluster expansions \cite{LinkC,LinkC1,LinkC3D}, and the functional renormalization group method \cite{FRGrev,FRG3D}. On the other hand, the numerical simulations are extremely challenging, and the finite-size algorithms, including exact diagonalization (ED), QMC and DMRG, suffer severe finite-size effects, which are quite consuming for large systems and can access infinite systems only by utilizing finite-size scaling.

    To treat the correlations in many body systems, one usually starts by evoking the ideas of ``mean field'', ``bath'' or boundary conditions. Analytical methods such as the Hartree-Fock mean-field theory and the saddle point approximation in path integral are commonly used. In fact, for lattice models the ``mean field'' idea goes back to the single-site Weiss method, applied first for classical magnetic models \cite{Weiss1907}. Contemporary mean field methods for lattice models include Guztwiller ansatz for bosons and/or fermions, or pairing approaches (Bogoliubov-like for bosons, or Bardeen-Cooper-Schrieffer-like for fermions) –- for an overview of these and other methods see Ref. \cite{Lewensteinbook} and references therein. In the context of the present work it is important to mention the ``cluster mean field theory'' (CMFT), where the mean field \`a la Weiss is combined with exact diagonalization on clusters (for recent developments see \cite{Yamamoto09,Ren14} and references therein). It is also worth mentioning ``entanglement mean field theory” (EMFT) \cite{SenDe12,SenDe12a}, which for spin models is formulated on few spin clusters, demanding self-consistency of entanglement properties. Both CMFT and EMFT are close to the standard MFT that can give quite accurate description of standard (Landau-Ginsburg-like) ordered and disordered phases, but typically only far from criticality.

    For thermal and open systems, one popular way is to introduce a ``heat bath'' to mimic the interactions between the system and the environment \cite{HeatBath}. Regarding numerical approaches, the density functional theory (DFT), also known as \textit{ab-initio} first-principle calculations \cite{DFTrev}, was built by extending the Thomas-Fermi approximation of homogeneous electron gas to the inhomogeneous electron system \cite{TFliquid}. Its huge success in condensed matter physics, quantum chemistry, and materials science largely relies on the simplicity and unification, \textit{``using a popular code, a standard basis, and a standard functional approximation''} \cite{DFTrev}.
    
    In order to handle strong correlations, several schemes were developed in the spirit of DFT. The examples include the dynamic mean-field theory \cite{DMFT11,DMFT12,DMFT13} that maps a lattice model (such as the Hubbard model) onto a quantum impurity model subject to self-consistent conditions, and the density matrix embedding theory (DMET) \cite{DMET} that was proposed aiming at a better consideration of the entanglement, thanks to the accompanying explosive advances in both quantum information science \cite{EntQinfo} and condensed matter physics \cite{EntCSC}. However, it is difficult to use these algorithms to study long-range ordered states or phase transitions. To probe the disordered ground-states (e.g. the spin liquids in the infinite frustrated systems), it was proposed to signal the disordered nature by simulating a finite system with random boundary conditions \cite{RandomBound}.

    \begin{figure}[tbp]
    	\includegraphics[angle=0,width=1\linewidth]{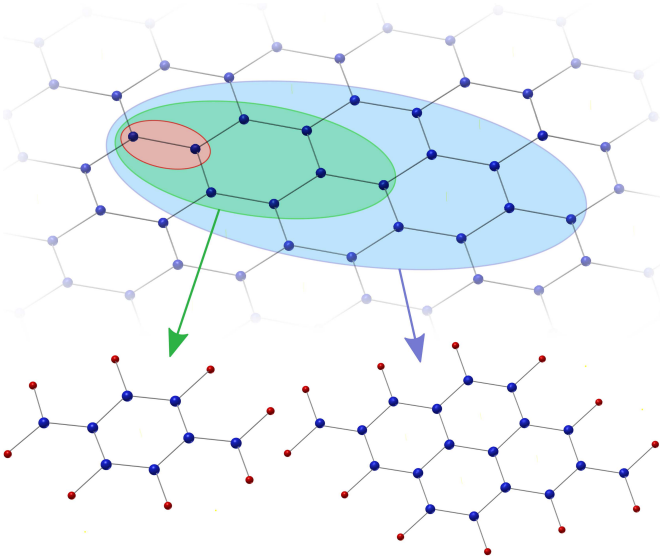}
    	\caption{(Color online) The system on an infinite lattice is transformed into one defined on finite clusters embedded in the entanglement bath. We take the $(8+8)$-site and $(18+12)$-site clusters for the simulations on honeycomb lattice, where the first contains 8 physical (blue balls) and 8 bath sites (red balls), and the second contains 18 physical and 12 bath sites. The entanglement bath is calculated by choosing two sites as the supercell (small circle). The legs stand for the interactions between the connecting sites.}
    	\label{fig-HCcluster}
    \end{figure}

	\textbf{Our proposal: mimicking many-body systems by few-body ones.}
	In general, as illustrated in Fig. \ref{fig-HCcluster}, the central idea of our work is to optimally find the few-body Hamiltonian to mimic the infinite model, without any prior knowledge of the ground state. The few-body model contains the physical sites in a finite cluster and the ``bath'' sites around it. The few-body Hamiltonian consists of two parts: the interactions among the physical sites (blue balls) within the cluster, and those [Eq. (\ref{eq-Hbath})] between the boundary physical sites and the bath sites (red balls). The physical-bath interactions are represented by some local Hamiltonians, which reproduce the quantum entanglement between the cluster and the bath, in such a way that the many-body effects from the infinite environment are well captured in the few-body simulations. Then the ground-state information of the infinite system is encoded in the reduced density matrix of the few-body ground state after tracing the bath degrees of freedom.
	
	The theoretical scheme we propose is a higher-dimensional generalization of the \textit{ab-initio} optimization principle (AOP) formulated with TN \cite{AOP}, and originally developed for infinite 1D systems with translational invariance. The idea is to find the simplest eigenvalue equations that encodes the infinite TN contraction problem. Besides its simplicity in the implementation, the 1D AOP has proved to have several computational advantages over other established algorithms, such as iTEBD and iDMRG \cite{DMRG}. For the purpose of the present work, the main advantage of the AOP is its flexibility and implications in high dimensions: without any substantial conceptual changes, the AOP can be readily extended to 2D and even 3D systems with high efficiency. Furthermore, the dynamic correlation length and the first excitation gap can be straightforwardly extracted.
	
	Our scheme consists of two stages: (1) compute physical-bath Hamiltonian and (2) solve the few-body Hamiltonian. In the first stage, by choosing the dimension $D$ of the bath site and a supercell that obeys the translational invariance, we start from the original Hamiltonian of the system and construct a set of self-consistent eigenvalue equations. Their solution gives the Hamiltonian $\mathcal{\hat{H}}^{\partial}$ [Eq. (\ref{eq-Hbath})] that describes the interactions between a physical and a bath sites. Such equations in fact encode an optimal zero-loop TN approximation of the state. This approximation directly enters the thermodynamic limit with a Bethe TN state ansatz \cite{TTN1,TTN2}, and already gives us the first glance of the ground state with good accuracy especially for the gapped states \cite{SimpleUpdate,ODTNS,NCD,SimpleTN3D,TTN1,TTN2}.
	
	The aim of the second stage is to construct the few-body Hamiltonian $\mathcal{\hat{H}}^{FB}$, and solve its ground state $|\tilde{\Phi} \rangle$ by e.g., DMRG (with certain dimension cut-off's denoted by $\chi$). $\mathcal{\hat{H}}^{FB}$ is formed by all the physical interactions inside a chosen cluster and several physical-bath interactions given by $\mathcal{\hat{H}}^{\partial}$. The choice of the cluster is very flexible. The ground-state properties of the infinite system is then encoded in the ground state  $|\tilde{\Phi} \rangle$  of $\mathcal{\hat{H}}^{FB}$. In other words, quantities such as energy, magnetization and entanglement of the infinite ground state are obtained from the density matrix of  $|\tilde{\Phi} \rangle$ by tracing all the bath degrees of freedom [Eq. (\ref{eq-RDM})].
	
	% Our numerical results show that the Heisenberg model on infinite honeycomb lattice is accurately simulated by a $\mathcal{\hat{H}}^{FB}$ that only contains $N_p=18$ physical sites surrounded by $N_b=12$ bath sites. The discrepancy compared with the state-of-the-art TN algorithm is $\mathcal{O}(10^{-3})$. For the 3D Heisenberg models on infinite simple cubic lattice, the ground-state properties and the critical behaviors near the quantum phase transition point are faithfully captured with only $N_p=8$ physical and $N_b=24$ bath sites.
	
	Our numerical results show, for instance, that the Heisenberg model on infinite honeycomb lattice is accurately simulated by a $\mathcal{\hat{H}}^{FB}$ that only contains $N_p=18$ physical sites surrounded by $N_b=12$ bath sites. For the 3D Heisenberg models on infinite simple cubic lattice, the ground-state properties including the critical behaviors near the quantum phase transition point are faithfully captured with only $N_p=8$ physical and $N_b=24$ bath sites. The discrepancy (such as energy) compared with the state-of-the-art TN algorithm is around $\mathcal{O}(10^{-3})$.
	
	The algorithm built from our scheme possesses several advantages (see Appendix G). The algorithm can directly reaches the thermodynamic limit by means of the physical-bath interactions on the boundary, thus has no conventional finite-size effects compared with the finite algorithms such as ED and DMRG. The strongly-correlated effects of the infinite models are accurately considered, and the many-body features, e.g., entanglement and criticality, can be efficiently captured, thus our scheme goes beyond the mean-field-based methods such as DFT \cite{DFTrev} and DMFT \cite{DMFT11,DMFT12,DMFT13}. Comparing with DMFT where the original model is approximated by an impurity model in a bath, we approximate the infinite-size system into a few-body model that contains its original interactions and the emergent physical-bath interactions. The accuracy is enhanced by fully considering all interactions in the cluster, thus outperforms the Bethe TN-based algorithms \cite{SimpleUpdate,ODTNS,NCD,SimpleTN3D,TTN1,TTN2}. In higher dimensions, the computational cost of our scheme is much lower than, e.g., the TN renormalization group algorithms \cite{TRG,MERA,PEPS11,PEPS11,CTMRG0,CTMRG11,CTMRG12,CTMRG3D}. It has no sign problem \cite{QMCsign} thus can be used to simulate frustrated and fermionic systems.
	
	The construction of $\mathcal{\hat{H}}^{FB}$ makes it possible to investigate the many-body effects in experiments by designing the few-body models --- quantum simulators described by the predicted Hamiltonians. The many-body behaviors are expected to be observed in the bulk of the few-body model. The feasibility of realizing $\mathcal{\hat{H}}^{FB}$ in cold atom experiments is supported by several facts observed in our numerical simulations: the few-body Hamiltonian has the same interaction length as the original Hamiltonian; with a proper tolerance of error, say $\mathcal{O}(10^{-2})$, the size of the few-body model can be very small. Especially for spin-$1/2$ models on simple cubic lattice, we show that it is sufficient to use only the spin-$1/2$'s as the bath sites. The few-body Hamiltonian then is just a small spin-$1/2$ system that includes some special interactions (given by $\mathcal{\hat{H}}^{\partial}$) on the boundary.
	
	% Then, we calculate  with {\B{More interestingly, the bath degrees of freedom can simply be given by higher spins, and the physical-bath interactions are local. Thus, the few-body Hamiltonian [see Eq. (\ref{eq-Hfewbody})] can be realized in, e.g., cold atom experiments.}}
	
	%{\B{Numerically speaking, our scheme gives birth to a novel TN-based algorithm that is able to efficiently access 2D and 3D infinite many-body models with no sign problem.}} It can be reduced to several established methods and beats them regarding both the accuracy and efficiency . The entanglement bath is calculated on an optimal tree \cite{TTN1,TTN2,SimpleUpdate,ODTNS,NCD} to guarantee the high efficiency and good accuracy for the bath calculations of higher-dimensional systems. Meanwhile, compared with ED or finite DMRG, the finite-size effects are reduced by the entanglement bath, and the lost information due to the tree approximation compared with the tree approximation \cite{NCD} or simple update \cite{SimpleUpdate} is retrieved by considering all the couplings in the chosen cluster (see Fig. \ref{fig-HCmag} and related discussion).
	
	% We benchmark our approach by calculating the ground-state energy of the Heisenberg model on honeycomb lattice. Then we apply our approach to the 3D simple cubic lattice, where the ground-state properties of Heisenberg anti-ferromagnet and the quantum phase transition of transverse Ising model are investigated. {\B{The experimental realization of the few-body Hamiltonian is also discussed.}}

\begin{figure}[tbp]
	\includegraphics[angle=0,width=0.85\linewidth]{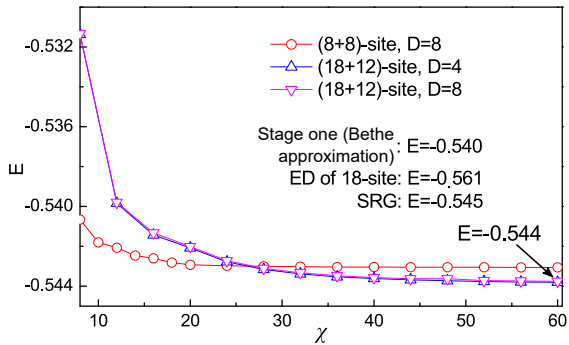}
	\caption{(Color online) The ground-state energy $E$ (per site) of the Heisenberg model on honeycomb lattice. The cluster we choose is ($N_p+N_b$)-site where $N_p$ and $N_b$ denote the number of physical and bath sites, respectively (see Fig. \ref{fig-HCcluster}). The ED on the 18-site cluster with periodic boundary condition suffers severe finite-size effects, and the tree approximation (simply from the bath calculations) underestimates long-range correlations. Our results are consistent with second renormalization group (SRG) of TN, showing that both finite-size effects and the error from the tree approximation are largely reduced.}
	\label{fig-E0HC}
\end{figure}

\section{Numerical results}

    \textbf{Heisenberg model on honeycomb lattice.}
    We simulate the ground-state properties of the Heisenberg model on honeycomb lattice, which is on a gapless point and considered to be challenging to simulate. The Hamiltonian is the summation of the two-body interactions as
    \begin{eqnarray}
    %\hat{H} = \sum_{\langle i,j \rangle} (J_x \hat{S}_i^x \hat{S}_i^x + J_y \hat{S}_i^y \hat{S}_i^y + J_z \hat{S}_i^z \hat{S}_i^z),
    \hat{H} = \sum_{\langle i,j \rangle} \hat{H}(i,j).
    \label{eq-PhysicalH}
    \end{eqnarray}
    For Heisenberg model, we have $\hat{H}(i,j)=J_x \hat{S}^x(i) \hat{S}^x(j) + J_y \hat{S}^y(i) \hat{S}^y(j) + J_z \hat{S}^z(i) \hat{S}^z(j)$, with $\hat{S}^{\alpha}(i)$ ($\alpha = x,y,z$) the $\alpha$ component of the spin-1/2 operators on the $i$-th site and $J_{\alpha}$ the coupling constants.

    In stage one, the bath is calculated by choosing two neighboring sites as the supercell. It means $\mathcal{\hat{H}}^{FB}$ that appears in this stage contains $N_p=2$ physical and $N_b=4$ bath sites. For stage two, we choose two different clusters to construct $\mathcal{\hat{H}}^{FB}$, which contains $N_p=8$ physical sites surrounded by $N_b=8$ bath sites and $N_p=18$ physical sites with $N_b=12$ bath sites, respectively (Fig. \ref{fig-HCcluster}). We utilize finite DMRG \cite{DMRG} to solve the ground state of $\mathcal{\hat{H}}^{FB}$.
	
	The ground-state energy $E$ with different dimensions of the bath site $D=4$ and $8$ is shown in Fig. \ref{fig-E0HC}. One can see that $E$ converges rapidly by increasing the dimension cut-off of DMRG $\chi$ to $E=-0.543$ and $-0.544$ for the two clusters, respectively. With larger $D$, the bath will be able to carry more entanglement and lead to a better accuracy.
	The accuracy will also be improved by increasing $\chi$ since the result will approach to the exact ground state of $\mathcal{\hat{H}}^{FB}$ with no DMRG error. When $\chi$ is sufficiently large, the errors inside the cluster due to the tree approximation, Trotter discretization and truncations will vanish.
	
	For a comparison, the ground-state energy by ED on such a cluster of 18 spins (Fig. \ref{fig-HCcluster}) with periodic boundary condition is $E=-0.561$, which suffers severe finite-size effects. The result solely by bath calculation (tree approximation) is $E=-0.540$, and by second renormalization group (SRG) \cite{SRG} of TN is $E=-0.545$. SRG belongs to the state-of-the-art TN approaches for simulating 2D ground states with a high accuracy. The difference compared with our results are only $\mathcal{O}(10^{-3})$.
	
	\begin{figure}[tbp]
		\includegraphics[angle=0,width=1\linewidth]{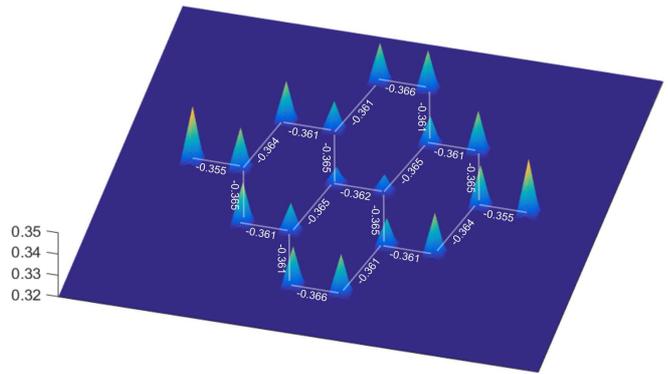}
		\caption{(Color online) ``Finite-size effects'' of AOP from the ground-state magnetization (absolute value) and the bond energies $e=\langle \hat{S}_i \hat{S}_j \rangle$ of the Heisenberg model on honeycomb lattice. Here we the cluster with $N_p=18$ physical and $N_b=12$ bath sites. Each peak shows the absolute values of the local magnetization of the physical sites, which ranges from $M=0.329$ (center) to $0.347$ (boundary). We take $D=8$ and $\chi=60$. For comparison, the results from the tree approximation in the first stage are $e=-0.360$ and $M=0.347$, and those from SRG are $e=-0.363$ and $M=0.310$.}
		\label{fig-HCmag}
	\end{figure}
	
	To further investigate the effects of the finiteness of the clusters, we calculate the nearest-neighbor bond energies $e=\langle \hat{S}_i \hat{S}_j \rangle$ and magnetizations with the cluster of $N_p=18$ and $N_b=12$ (Fig. \ref{fig-HCmag}). The changes of both quantities in different positions of the cluster is mostly $\mathcal{O}(10^{-2})$. By comparing with the tree results in the bath calculation and SRG, we find that the bond energies and magnetization on the boundary of the cluster are very close to the tree results, and in the middle where the ``boundary effects'' as well as the difference between our results and the SRG are minimal.
	
	Our simulations show that without increasing the computational cost much, the finite-size effects are suppressed by introducing the entanglement bath, and at the same time the error from the tree approximation are reduced by choosing larger clusters.
	
	% ---------------------------------------------------
	
	\begin{figure}[tbp]
		\includegraphics[angle=0,width=1\linewidth]{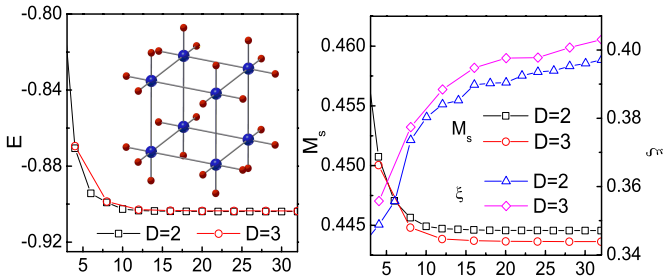}
		\caption{(Color online) The ground-state energy $E$ of the Heisenberg model on simple cubic lattice versus $\chi$ is shown in the left figure. The simulation by QMC on a ($10\times 10\times 10$) lattice with periodic boundary condition gives $E=-0.902$. The inset shows the cluster with $N_p=8$ physical (blue balls) and $N_b=24$ bath sites (red balls) used in the second stage in our AOP approach for the simulations on cubic lattice. The legs stand for the interactions between the connecting sites. In the right one, we show the staggered magnetization $M_s$ and correlation length $\xi$ of the ground state. We take $D=2$ and $3$.}
		\label{fig-CubicEg}
	\end{figure}
	
	\textbf{Spin models on simple cubic lattice.}
	We investigate the ground-state properties and quantum phase transitions in the spin models on simple cubic lattice. For bath calculations, the supercell is chosen to be two neighboring sites, giving a $\mathcal{\hat{H}}^{FB}$ with $N_p=2$ physical and $N_b=10$ bath sites. In stage two, we choose a cubic with $N_p=8$ physical and $N_b=24$ bath sites to construct $\mathcal{\hat{H}}^{FB}$ (inset of Fig. \ref{fig-CubicEg}).
	
	The ground-state energy $E$, staggered magnetization $M_s$ and dynamic correlation length $\xi$ of the antiferromagnetic Heisenberg model on simple cubic lattice are shown in Fig. \ref{fig-CubicEg}. The energy converges to $E=-0.904$, while that from QMC \cite{QMCworm,QMChuang,Huang} on a ($10\times 10\times 10$) lattice with periodic boundary condition is $E=-0.902$. Note the result from the tree approximation in the first stage is $E=-0.892$, which is already quite accurate. For $M_s$ and $\xi$, the finite-size effects are much stronger for our QMC calculations. The AOP simulations show that $M_u=0$ (uniform magnetization), $M_s=0.445$ and $\xi=0.405$. Our results are consistent with the widely accepted consensus, that its ground state is an antiferromagnetic ordered (N\'eel) state with a short correlation length.
	
	\begin{figure}[tbp]
		\includegraphics[angle=0,width=1\linewidth]{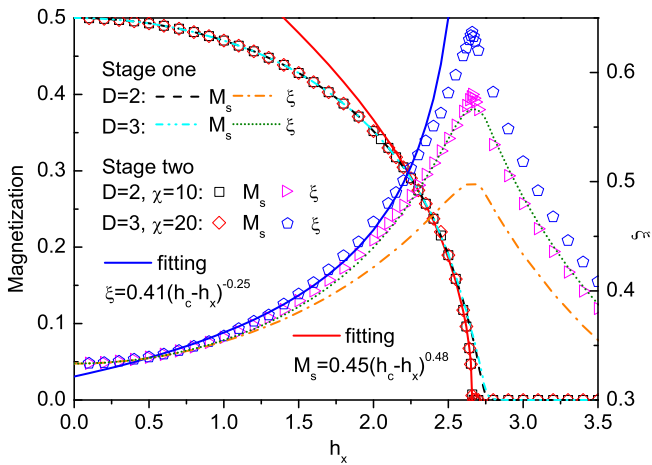}
		\caption{(Color online) Staggered magnetization per site $M_s$ and dynamic correlation length $\xi$ of the ground state of the transverse Ising model on simple cubic lattice. The results obtained in Stage one (Bethe approximation) and Stage two are shown for comparison. We take $D=2$, $\chi=10$, and $D=3$, $\chi=20$. The quantum phase transition is found to occur at $h_c = 2.66$. The critical behaviors are obtained by fitting the data from the results of Stage two near $h_c$, where we have $M_{s} \propto (h_c-h_x)^{0.48}$ and $\xi \propto (h_c-h_x)^{-0.25}$.}
		\label{fig-CubicMag}
	\end{figure}
	
	We investigate the quantum phase transition of the anti-ferromagnetic Ising model in a transverse field on simple cubic lattice (Fig. \ref{fig-CubicMag}), where the Hamiltonian reads $\hat{H}= \sum_{<i,j>} \hat{S}^x(i) \hat{S}^x(j)  + h \sum_i \hat{S}^z(i)$. For a comparison, we try different dimension cut-offs with $D=2$, $\chi=10$, and $D=3$, $\chi=20$. The critical field is found to be around $h_c=2.66$, consistent with the results from other algorithms (Table \ref{tab-Fields}).
	
	\begin{table}[tbp]
		\caption{The values of the critical field $h_{c}$ with the perturbation expansions (PE) \cite{CubicIsing}, cluster quantum Monte Carlo (cQMC) \cite{QIsing3DQMC}, linked-cluster expansions (LCE) \cite{LinkC3D}, mean field theory (MFT) \cite{QIP&TinTIS}, and our AOP simulations in Stages one and two.}
		\begin{tabular*}{8cm}{@{\extracolsep{\fill}}lcccccc}
            \\
			\hline\hline
			& PE & cQMC & LCE & MFT & Stage one & Stage two   \\ \hline
			$h_{c}$ & 2.60 & 2.58 & 2.65 & 3 & 2.8 & 2.66
			\\ \hline\hline
			\label{tab-Fields}
		\end{tabular*}
	\end{table}
	
	Our results show that from the few-body Hamiltonian, the scaling behavior in the critical region can be faithfully captured and the critical exponents are consistent with the results obtained by other methods. Meanwhile, obvious improvement in Stage two is observed compared with Stage one (Bethe approximation with a small cluster). By fitting the data in Stage two of $D=3$ and $\chi=20$ near the critical point, we find
	\begin{eqnarray}
	M_s \propto (h_c-h_x)^{\beta^{\ast}},
	\end{eqnarray}
	with $\beta^{\ast} = 0.48$, which is close to but slightly larger than the perturbation expansions result $\beta^{\ast} = 0.46$. Note that the exponent from the fitting on the data of Stage one is not reliable, which gives $\beta^{\ast} = 1$.
	
	We also calculate the dynamic correlation length $\xi$, which shows a peak at the critical point and scales as
	\begin{eqnarray}
	\xi \propto (h_c-h_x)^{-\sigma},
	\end{eqnarray}
	with $\sigma=0.25$ near the critical point in Stage two. In Stage one, we have $\sigma=0.23$. The exponent of the (spatial) correlation length by the perturbation expansions is $\sigma=0.5$ \cite{CubicIsing} . The discrepancy might be caused by the errors from both sides. Regarding the TN algorithms, the correlation length in the critical region will diverge with the scaling of the bath dimension $D$ as well as the DMRG dimension cut-off $\chi$ (unlike $M_s$ which converges to zero). Thus, it is difficult to directly extract the exponent of $\xi$ with fixed dimensions. The good thing is that the algebraic behavior of $\xi$ is clearly observed. What is open is how to get an accurate value of $\sigma$ by the scaling factors versus not only $h_x$ but also $\chi$ and $D$. See more discussions about the error of correlation functions with TN approaches in Appendix F.
	
	% ===================================================
	
\section{Method: higher-dimensional \textit{ab-initio} optimization principle approach}
	
	The idea of AOP scheme \cite{AOP} is, without any previous knowledge of the ground state, to transfer the infinite system to a finite one embedded in an entanglement bath. In the language of TN, the idea is to encode the contraction of an infinite TN in a simplest-possible local function that can be exactly computed, with smallest-possible number of inputs. The 1D version is briefly presented in Appendix A, where we also explain the relations and differences compared with the AOP approach in higher dimensions. To present the approach in high dimensions, we take the 2D spin model with nearest-neighbor couplings on honeycomb lattice as an example. The implementation can be easily generalized to other models on 2D and 3D lattices.
	
	\begin{figure}[tbp]
		\includegraphics[angle=0,width=1\linewidth]{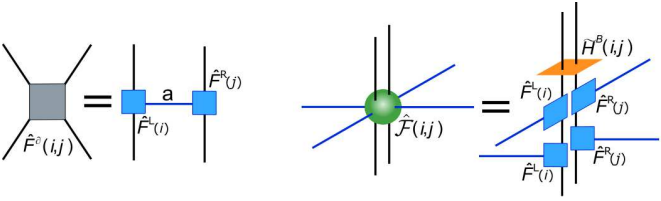}
		\caption{(Color online) The left figure shows Eq. (\ref{eq-Fsvd}). The right one shows the construction of the cell tensor given by Eq. (\ref{eq-getF}).}
		\label{fig-cellT_honeycomb}
	\end{figure}
	
	\textbf{Stage One: calculate the entanglement bath.} The first stage is to calculate the entanglement bath represented by a set of tensors dubbed as \textit{boundary tensors}. They are obtained by solving a set of self-consistent eigenvalue equations [see Eqs. (\ref{eq-effectH})-(\ref{eq-M4}) below]. These equations are parametrized by the Hamiltonian as well as by the boundary tensors themselves, thus they can be solved in an alternative way: starting from an arbitrary guess, we update one boundary tensor by fixing all others as the parameters of the equations, and iterate such a procedure for every tensor until the fixed point is reached.
	
	Though our method is based on the TN representation of the imaginary-time evolution with Trotter-Suzuki decomposition \cite{Trotter} like several existing methods \cite{iTEBD,PEPS11,PEPS12,CTMRG0,CTMRG11,CTMRG12}, the idea here is to encode the TN in the eigenvalue equations \cite{NoteEig,AOP} instead of contracting the TN. On the other hand, the implementation in this stage is borrowed from the generalization of DMRG on an infinite tree \cite{TTN1,TTN2}, which can be easily extended to 3D models with high efficiency. In the DMRG language, the (convergent) boundary tensors can be understood as the infinite environment of the tree brunches.
	
	To begin with, one chooses a supercell that obeys the translational invariance, e.g. two sites connected by a parallel bond (see the smallest shaded circle in Fig. \ref{fig-HCcluster}), and construct the \textit{cell tensor} that parametrizes the eigenvalue equations. The bulk interaction is simply the coupling between these two spins, i.e. $\hat{H}^B(i,j) = \hat{H}(i,j)$. For the interactions between two neighboring supercells, we define the two-body operator $\hat{F}^{\partial}(i,j) = I-\tau \hat{H}(i,j)$ and do the singular value decomposition (Fig. \ref{fig-cellT_honeycomb}) as
	\begin{eqnarray}
	\hat{F}^{\partial}(i,j) = \sum_{a} \hat{F}^{L}(i)_a \otimes \hat{F}^{R}(j)_a.
	\label{eq-Fsvd}
	\end{eqnarray}
	We dub $a$ as the \textit{boundary index}.
	
	To obtain the TN of the imaginary-time evolution, we define the \textit{cell tensor} that is the product of the (shifted) bulk Hamiltonian with $\hat{F}^{L}_a$ and $\hat{F}^{R}_a$ (Fig. \ref{fig-cellT_honeycomb}) as
	\begin{eqnarray}
	\hat{\mathcal{F}}(i,j)_{a_1a_2a_3a_4}= \hat{F}^{L}(i)_{a_1} \hat{F}^{L}(i)_{a_2} \hat{F}^{R}(j)_{a_3} \hat{F}^{R}(j)_{a_4} \tilde{H}^B(i,j),
	\label{eq-getF}
	\end{eqnarray}
	with $\tilde{H}^B(i,j) = I- \tau \hat{H}^B(i,j)$. Note that both $\hat{F}^{\partial}(i,j)$ and $\tilde{H}^B(i,j)$ can be different from the current choice. $\hat{F}^{\partial}(i,j)$ can be different when the model has long-range interactions, and $\tilde{H}^B(i,j)$ can be different when choosing different subsystems to define the supercell. $\hat{\mathcal{F}}(i,j)$ \cite{NoteIndex} can be understood as a set of quantum operators defined in the Hilbert space of the supercell (spins $i$ and $j$) labeled by the boundary indexes $a_1$, $a_2$, $a_3$ and $a_4$. Similar to 1D AOP \cite{AOP}, $\tau$ is in fact the Trotter step, and $\hat{\mathcal{F}}(i,j)$ gives the TN representation of $I-\tau \hat{H}$ with an error $\mathcal{O}(\tau^2)$.
	
	\begin{figure}[tbp]
		\includegraphics[angle=0,width=1\linewidth]{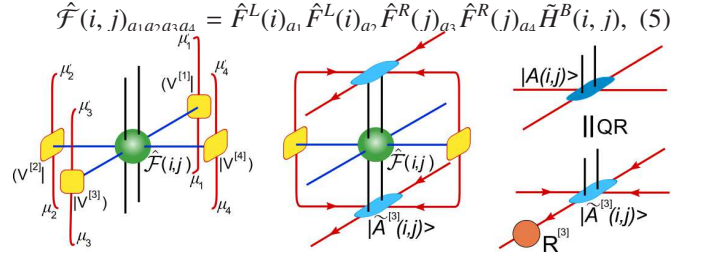}
		\caption{(Color online) The graphic representations of $\hat{\mathcal{H}}(i,j)$ in Eq. (\ref{eq-effectH}) and $M^{[3]}$ in Eq. (\ref{eq-M3}) are given in the left and middle figures, respectively. The QR decomposition [Eq. (\ref{eq-QR})] of the central tensor $|A(i,j)\rangle$ is shown in the right figure, where the arrows indicate the orthogonality of $|\tilde{A}^{[3]}(i,j)\rangle$ [Eq. (\ref{eq-OrtA3})].}
		\label{fig-HoneycombEigEqs}
	\end{figure}
	
	Then, with the boundary tensors $|V^{[x]})$ (guessed or previously obtained in the last iteration) and the cell tensor $\hat{\mathcal{F}}(i,j)$, we define five eigenvalue equations as
\begin{widetext}
	\begin{eqnarray}
		\hat{\mathcal{H}}(i,j)_{\mu_1'\mu_2'\mu_3'\mu_4',\mu_1\mu_2\mu_3\mu_4} = \sum_{a_1a_2a_3a_4} (V^{[1]}|_{a_1\mu_1 \mu_1'} (V^{[2]}|_{a_2\mu_2 \mu_2'} \hat{\mathcal{F}}(i,j)_{a_1a_2a_3a_4} |V^{[3]})_{a_3\mu_3 \mu_3'} |V^{[4]})_{a_4\mu_4 \mu_4'}, \label{eq-effectH}\\
		M^{[1]}_{a_1\mu_1\mu_1',a_3\mu_3\mu_3'} = \sum_{a_2a_4\mu_2\mu_2'\mu_4\mu_4'}(V^{[2]}|_{a_2\mu_2 \mu_2'} \langle \tilde{A}^{[1]}(i,j)|_{\mu_1'\mu_2'\mu_3'\mu_4'} \hat{\mathcal{F}}(i,j)_{a_1a_2a_3a_4} |\tilde{A}^{[1]}(i,j)\rangle_{\mu_1\mu_2\mu_3\mu_4} |V^{[4]})_{a_4\mu_4 \mu_4'},\label{eq-M1}\\
		M^{[2]}_{a_2\mu_2\mu_2',a_4\mu_4\mu_4'} = \sum_{a_1a_3\mu_1\mu_1'\mu_3\mu_3'} (V^{[1]}|_{a_1\mu_1 \mu_1'} \langle \tilde{A}^{[2]}(i,j)|_{\mu_1'\mu_2'\mu_3'\mu_4'} \hat{\mathcal{F}}(i,j)_{a_1a_2a_3a_4} |\tilde{A}^{[2]}(i,j)\rangle_{\mu_1\mu_2\mu_3\mu_4} |V^{[3]})_{a_3\mu_3 \mu_3'},\label{eq-M2}\\
		M^{[3]}_{a_1\mu_1\mu_1',a_3\mu_3\mu_3'} = \sum_{a_2a_4\mu_2\mu_2'\mu_4\mu_4'} (V^{[2]}|_{a_2\mu_2 \mu_2'} \langle \tilde{A}^{[3]}(i,j)|_{\mu_1'\mu_2'\mu_3'\mu_4'} \hat{\mathcal{F}}(i,j)_{a_1a_2a_3a_4} |\tilde{A}^{[3]}(i,j)\rangle_{\mu_1\mu_2\mu_3\mu_4} |V^{[4]})_{a_4\mu_4 \mu_4'},\label{eq-M3}\\
		M^{[4]}_{a_2\mu_2\mu_2',a_4\mu_4\mu_4'} = \sum_{a_1a_3\mu_1\mu_1'\mu_3\mu_3'} (V^{[1]}|_{a_1\mu_1 \mu_1'} \langle \tilde{A}^{[4]}(i,j)|_{\mu_1'\mu_2'\mu_3'\mu_4'} \hat{\mathcal{F}}(i,j)_{a_1a_2a_3a_4} |\tilde{A}^{[4]}(i,j)\rangle_{\mu_1\mu_2\mu_3\mu_4} |V^{[3]})_{a_3\mu_3 \mu_3'}.\label{eq-M4}
	\end{eqnarray}
\end{widetext}
	
	By solving the leading eigenvector of $\hat{\mathcal{H}}(i,j)$ given by Eq. (\ref{eq-effectH}), we obtain a tensor $|A(i,j)\rangle_{\mu_1\mu_2\mu_3\mu_4}$ dubbed as \textit{central tensor} with $\mu_1$, $\mu_2$, $\mu_3$ and $\mu_4$ called \textit{virtual indexes} according to the TN terminology. The central tensor can be considered as a state in the Hilbert space of the supercell labeled by four virtual indexes.
	
	Meanwhile, $|V^{[x]})$ is obtained as the (left) leading eigenvector of $M^{[x]}$ [Eqs. (\ref{eq-M1})-(\ref{eq-M4})]. One can see that $M^{[x]}$ is defined by the isometries $|\tilde{A}^{[x]}(i,j) \rangle$ that is obtained by the QR decomposition of $|A(i,j)\rangle$ (referring to the $x$-th virtual bond $\mu_x$) of the central tensor. For example for $x=3$, we have (Fig. \ref{fig-HoneycombEigEqs})
	\begin{eqnarray}
	|A(i,j)\rangle_{\mu_1\mu_2\mu_3\mu_4} = \sum_{\nu} |\tilde{A}^{[3]}(i,j)\rangle_{\mu_1 \mu_2 \nu \mu_4} R^{[3]}_{\nu \mu_3}.
	\label{eq-QR}
	\end{eqnarray}
	$|\tilde{A}^{[3]}(i,j)\rangle$ is orthogonal, satisfying
	\begin{eqnarray}
	\sum_{\mu_1 \mu_2 \mu_4} \langle \tilde{A}^{[3]}(i,j)|_{\mu_1 \mu_2 \mu_3 \mu_4} |\tilde{A}^{[3]}(i,j)\rangle_{\mu_1 \mu_2 \mu_3' \mu_4} = I_{\mu_3 \mu_3'}.
	\label{eq-OrtA3}
	\end{eqnarray}
	
	These isometries play the role of the renormalization group flow in the tree DMRG \cite{TTN1,TTN2}. Similarly, $|V^{[x]})_{a_x\mu_x \mu_x'}$ can be understood as a ``state'' defined in the space of the boundary index $a_x$ labeled by $\mu_x$ and $\mu_x'$ \cite{NoteBracket}. The graphic representations of $\hat{\mathcal{H}}(i,j)$ and $M^{[3]}$ are given in Fig. \ref{fig-HoneycombEigEqs} as examples.
	
	One can see that these equations are parametrized by the solutions of others, and can be solved in an alternative way in practice. One can start with four random $|V^{[x]})$'s and calculate $|A(i,j)\rangle$ by solving the leading eigenvector of Eq. (\ref{eq-effectH}). Then one obtains $|\tilde{A}^{[x]}(i,j) \rangle$'s using Eq. (\ref{eq-QR}) and update the $|V^{[x]})$'s according to Eqs. (\ref{eq-M1})-(\ref{eq-M4}). Repeat this process until the central tensor and all boundary tensors converge.
	
	In fact, the ground-state properties can already be well extracted by the central tensor $|A(i,j)\rangle$. For example, the reduced density matrix of the supercell $\hat{\rho}(i,j) = \text{Tr}_{/(i,j)} |\Psi \rangle \langle \Psi |$ (with $|\Psi \rangle$ denoting the ground state of the infinite model) is well approximated by the central tensor as
	\begin{eqnarray}
	\hat{\rho}(i,j) \simeq \sum_{\mu_1\mu_2\mu_3\mu_4} |A(i,j)\rangle_{\mu_1\mu_2\mu_3\mu_4} \langle A(i,j)|_{\mu_1\mu_2\mu_3\mu_4}.
	\label{eq-rhoBath}
	\end{eqnarray}
	
	Since each boundary tensor can be understood as the environment of an infinite tree branch, the original model is actually approximated at this stage by one defined on an infinite tree. Note that when only looking at the tree locally (from one site and its nearest neighbors), it looks the same to the original lattice. Thus, the loss of information is mainly long-range, i.e., from the destruction of loops. Though it has been shown numerically by many previous work that the tree approximation is very accurate especially for gapped systems \cite{SimpleUpdate,ODTNS,NCD,SimpleTN3D}, we are still facing the difficulty of controlling the effects (errors) brought by such an approximation. More discussions about such a tree approximation are given in the Appendix B, starting from the state ansatz behind our approach. One can also find more details in the forth section of a recent review \cite{TNrev}. To further improve the precision in a systematic way, the next stage is to embed a much larger subsystem in the entanglement bath.
	
	\begin{figure}[tbp]
		\includegraphics[angle=0,width=1\linewidth]{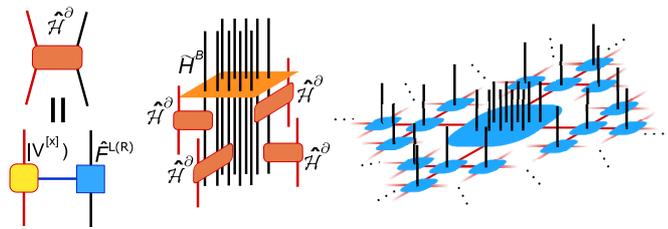}
		\caption{(Color online) The left figure shows the bath Hamiltonian $\hat{\mathcal{H}}^{\partial}$ [Eq. (\ref{eq-Hbath})] that gives the interaction between the corresponding physical and bath site. The few-body Hamiltonian in Eq. (\ref{eq-Hfinite}) is formed by the shifted bulk Hamiltonian and $\hat{\mathcal{H}}^{\partial}$ between every physical site on the boundary and a neighboring bath site. For simplicity, the middle figure only illustrates four of the $\hat{\mathcal{H}}^{\partial}$'s. The right one shows the ground-state ansatz of AOP approach, which is the bulk state of the few-body Hamiltonian entangled with several branches of infinite tree PEPS. In fact, the number of tree branches should equal to the number of the physical sites on the boundary (i.e. the number of $\hat{\mathcal{H}}^{\partial}$). For conciseness, we only illustrate four of the tree branches.}
		\label{fig-Hbath}
	\end{figure}
	
	\textbf{Stage two: construct the few-body Hamiltonian and solve it.} The second stage is to choose a finite cluster and use the obtained boundary tensors to construct a few-body Hamiltonian. All interactions inside the cluster will be fully considered to reduce the error from the tree approximation. The entanglement bath mimics the environment of the infinite tree branches, thus the algorithm directly accesses the thermodynamic limite and there is no conventional finite-size error that appears in, e.g. ED, DMRG or QMC.
	
	The embedding is based on the generalizations of $\hat{\mathcal{H}}(i,j)$ [Eq. (\ref{eq-effectH})] in stage one. From the formulation given above, one can see that $\hat{\mathcal{H}}(i,j)$ is actually the product of two parts. The first is the shifted Hamiltonian that contains all interactions inside the supercell (two neighboring sites in our example), and the second is in fact the physical-bath interactions (Fig. \ref{fig-Hbath}), whose Hamiltonian is written as
	\begin{eqnarray}
	\hat{\mathcal{H}}^{\partial}(n,x)_{\mu_x \mu_x'} = \sum_{a_x} \hat{F}^{L(R)}(n)_{a_x} |V^{[x]})_{a_x\mu_x\mu_x'}.
	\label{eq-Hbath}
	\end{eqnarray}
	
	Now we extend the supercell to a chosen larger cluster, where the few-body Hamiltonian denoted by $\mathcal{\hat{H}}^{FB}$ is written as
	\begin{eqnarray}
	\mathcal{\hat{H}}^{FB} = \prod_{\langle x \in cluster,n \in bath \rangle} \hat{\mathcal{H}}^{\partial}(n,x) \sum_{\langle i,j \rangle \in cluster} [I-\tau \hat{H}(i,j)].
	\label{eq-Hfinite}
	\end{eqnarray}
	Same as $\hat{\mathcal{H}}(i,j)$, $\mathcal{\hat{H}}^{FB}$ is also formed by two terms (Fig. \ref{fig-Hbath}). The first term is the product of several bath Hamiltonians that mimic the interactions between the cluster and the environment, and in the second term, the summation in Eq. (\ref{eq-Hfinite}) contains all couplings inside the cluster.
	
	The entanglement bath only ``interacts'' with the physical sites nearby according to the coupling distance of the original Hamiltonian. In our example with nearest-neighbor couplings, every physical sites on the boundary interact with a bath site, and thus, the number of $\hat{\mathcal{H}}^{\partial}(n,x)$ in the product above scales with the length of the boundary of the cluster. For this reason, $\mathcal{\hat{H}}^{FB}$ is the product/summation of sparse or local matrices, and its ground state can be efficiently solved by using the finite-size approaches, such as ED or DMRG.
	
	Note that if one takes the cluster as the supercell with two sites, Eq. (\ref{eq-Hfinite}) becomes exactly Eq. (\ref{eq-effectH}). The bath calculation itself can be considered as using ED to solve the $\mathcal{\hat{H}}^{FB}$ that contains only the supercell and the bath. The cluster can be arbitrarily chosen according to the computational capacity, and it does not have to obey the translational invariance of the model.
	
	With the ground state $|\tilde{\Phi}(i,j,\cdots) \rangle_{\{\mu\}}$ of $\mathcal{\hat{H}}^{FB}$, the physical properties such as energy, magnetization, etc., can be obtained from the density operator $\tilde{\rho}$ by tracing all degrees of freedom of the bath sites as
	\begin{eqnarray}
	\tilde{\rho}(i,j,\cdots) = \sum_{\{\mu\}} |\tilde{\Phi}(i,j,\cdots) \rangle_{\{\mu\}} \langle \tilde{\Phi}(i,j,\cdots) |_{\{\mu\}}.
	\label{eq-RDM}
	\end{eqnarray}
	Again, this is a generalization of Eq. (\ref{eq-rhoBath}).
	
\section{Discussions about experimental realizations}
	
	Our work provides a way of using few-body experiments to mimic many-body features of infinite systems. Since the few-body Hamiltonian only contains a handful of sites with local interactions, one could design cold-atom experiments to realize it in a lab. Specifically speaking in our examples, $\hat{\mathcal{H}}^{\partial}$ is the interaction between a physical spin and an artificial spin with $D$ (bath) degrees of freedom. Here, we assume that $\hat{\mathcal{H}}^{\partial}$ is Hermitian, which should be true due to the structure of the eigenvalue equations [Eqs. (\ref{eq-M1})-(\ref{eq-M4})] of the boundary tensors, where we have $|V^{[x]})_{a\mu\mu'} = |V^{[x]})_{a\mu'\mu}^{\ast}$. The task here is to get the coupling constants explicitly for implementing experiments.
	
	To this end, let us transform $\hat{\mathcal{H}}^{\partial}$ to the standard summation form. We define $\hat{H}^{\partial}$ that satisfies
	\begin{eqnarray}
	\hat{\mathcal{H}}^{\partial}(n,x) = I - \tau \hat{H}^{\partial}(n,x) + O(\tau^2).
	\label{eq-Hexp}
	\end{eqnarray}
	It means to the first order of $\tau$, $\hat{\mathcal{H}}^{\partial}$ is the evolution operator of a Hamiltonian $\hat{H}^{\partial}$ for an infinitesimal imaginary time. This relation is true because in Eq. (\ref{eq-Hbath}), $\hat{F}^{L(R)}$ is obtained by the decomposition of $I-\tau \hat{H}^B$, and the boundary tensor $|V^{[x]})$ has the similar structure since it forms an continuous MPS \cite{tMPS,cMPS} in the imaginary time direction.
	
	Then, the few-body Hamiltonian in Eq. (\ref{eq-Hfinite}) can be rewritten in a standard summation form as $\mathcal{\hat{H}}^{FB} = I - \tau \hat{H}^{FB} + O(\tau^2)$ with
	\begin{eqnarray}
	\hat{H}^{FB} = \sum_{\langle i,j \rangle \in cluster} \hat{H}(i,j) + \sum_{\langle x \in cluster,n \in bath \rangle} \hat{H}^{\partial}(n,x).
	\label{eq-Hfewbody}
	\end{eqnarray}
	The two summations contain the physical and physical-bath interactions, respectively, and all terms are local as discussed above. Again, $\mathcal{\hat{H}}^{FB}$ is the evolution operator of $\hat{H}^{FB}$ for an infinitesimal imaginary time to the first order of $\tau$, i.e. $\mathcal{\hat{H}}^{FB} \simeq e^{-\tau \hat{H}^{FB}}$.
	
	The coupling constants of the physical-bath interactions can be calculated by expanding $\hat{H}^{\partial}$ as
	\begin{eqnarray}
	\hat{H}^{\partial}(n,x) = \sum_{\alpha \alpha'} J_{\alpha \alpha'}(n,x) \mathcal{\hat{S}}^{\alpha'}(n) \hat{S}^{\alpha}(x).
	\label{eq-Expansion}
	\end{eqnarray}	
	with $J_{\alpha \alpha'}(n,x)$ the physical-bath coupling constants and $\{\hat{S}^{\alpha}\}$ and $\{\mathcal{\hat{S}}^{\alpha'}\}$ the corresponding spin operators (including identity) that give the complete basis for the Hermitian matrices. $\{\hat{S}^{\alpha}\}$ is in fact the physical spin operators. For $\{\mathcal{\hat{S}}^{\alpha'}\}$, one can generally choose the generators of SU(N) groups, which give a complete basis for an N-by-N Hermitian matrix. Then the bath spins should be SU(N) spins. If a symmetry \cite{DMRGsymme11,DMRGsymme12,Qspace} is used in the tensors, for example SU(2) symmetry for spin models, the bath spins are SU(2) spins with higher total momentum, and one will explicitly have the coefficients from the elements of $\hat{H}^{FB}$. Moreover, it is possible to translate the whole few-body Hamiltonian into the second-quantized picture, by expanding it with the bosonic or fermionic operators. The key is that the chosen operator basis have to completely expand the physical-bath Hamiltonian.
	
	From our numerical results, we can see that the properties of the infinite model can be accurately mimicked by very small bath dimension $D$ and cluster size. Suppose we set the tolerance of the experimental error as $\mathcal{O}(10^{-2})$. In this case, the cluster can be chosen as two sites. Then we have $N_p=2$, with $N_b=4$ for honeycomb lattice and $N_b=10$ for simple cubic lattice. For the spin-$1/2$ models on simple cubic lattice, the dimension of the bath sites can be chosen as $D=2$. This means the bath spins are simply spin-$1/2$, same as the physical ones, which makes it easy to implement in experiments.
	
	In short, the steps to mimic an infinite many-body system with a few-body model are as the following:
	\begin{itemize}
		\item Starting from the Hamiltonian of the target model [e.g. Eq. (\ref{eq-PhysicalH})], compute the physical-bath Hamiltonian $\mathcal{\hat{H}}^{\partial}$ [Eq. (\ref{eq-Hbath})] by our AOP algorithm.
		\item Write $\mathcal{\hat{H}}^{\partial}$ into $\hat{H}^{\partial}$ by Eq. (\ref{eq-Hexp}), so that the total Hamiltonian of the few-body model is in a standard summation form [Eq. (\ref{eq-Hfewbody})].
		\item According to the symmetry of the system, choose a set of matrix basis to expand $\hat{H}^{\partial}$ [Eq. (\ref{eq-Expansion})]. The basis will determine which kind of spins will be used as the bath sites, and the expansion coefficients will be the coupling constants.
		\item Build the few-body experiment with several physical sites in the bulk and bath sites on the boundary (e.g., Fig. \ref{fig-HCcluster} or the inset of Fig. \ref{fig-CubicEg}). The coupling constants in the bulk are the same as the target model, and the coupling constants on the boundary are given by the expansion coefficients of $\hat{H}^{\partial}$.
		\item Observe the properties of the bulk, which mimics the ground state of the infinite system.
	\end{itemize}
	
	% ===================================================
	
\section{Summary and outlook}
	
	We propose an \textit{ab-initio} TN approach that allows for accurate survey of the ground states of infinite many-body systems in higher dimensions by an effective few-body models embedded in an ``entanglement bath''. On one hand, our scheme gives to birth to an flexible and efficient numeric algorithm for quantum lattice models. Our approach can directly access the thermodynamic limit by introducing the physical-bath interactions, which outperforms the finite-size methods such as ED and DMRG. The embedding idea allows for efficient and accurate simulations of infinite 3D quantum models, surpassing the existing TN methods. It is free from the ``negative-sign'' problem and can access to frustrated spin and fermionic models. It can accurately capture many-body features including entanglement, phase transitions and critical behaviors, thus it goes beyond the DFT-based approaches. It could be readily applied to other $(d \le {3})$-dimensional systems and could be generalized to $(d \ge {4})$-dimensional models.
	
	In practice, our numerical simulations show that with only a handful of sites, the few-body models can accurately capture the many-body features of the infinite systems. With less than 18 physical and 12 bath sites, the difference between our results and the state-of-the-art TN methods is less than $\mathcal{O}(10^{-3})$. For the spin models on simple cubic lattice, the properties of the quantum phase transitions in a magnetic field, including the phase transition point and critical exponents, are faithful captured by the few-body model containing only 8 physical and 24 bath sites.
	
	On the other hand, the few-body Hamiltonian only contains local interactions among a handful of sites, it can be realized by, e.g., cold atoms or ions. It is possible to further improve the experiments by using the trick of synthetic gauge fields, where the higher spins, for instance, can be extended to lower spins in a synthetic dimension \cite{SynD}. We suggest to investigate infinite many-body systems by realizing the predicted few-body Hamiltonian with cold atoms or ions. The many-body phenomenon are expected to be observed in the bulk. Furthermore, our work exhibits a new perspective of designing novel quantum devices \cite{CSW} by utilizing the many-body properties that appears in the bulk of the few-body system, e.g., controlling the entanglement or quantum fluctuations by driving the system to approach or leave the critical region.

\section*{Acknowledgments}
	We acknowledge Leticia Tarruell, Ignacio Cirac, Emanuele Tirrito, Xi Chen, Nan Li and Jia Kong for enlightening discussions. CP appreciates ICFO (Spain) for the hospitality during her visit and is grateful to financial support from UCAS and ICFO. SJR is indebted to Nan Li for her kind help on illustrating the figures of lattices. CP thanks Yun-Hai Lan for modifying references format. This work was supported by ERC AdG OSYRIS (ERC-2013-AdG Grant No. 339106), the Spanish MINECO grants FOQUS (FIS2013-46768-P), FISICATEAMO (FIS2016-79508-P), and ``Severo Ochoa'' Programme (SEV-2015-0522), Catalan AGAUR SGR 874, Fundaci\'o Cellex, EU FETPRO QUIC, EQuaM (FP7/2007-2013 Grant No. 323714), and CERCA Programme / Generalitat de Catalunya. SJR acknowledges Fundaci\'o Catalunya - La Pedrera $\cdot$ Ignacio Cirac Program Chair. GS and CP were supported by the MOST of China (Grant No. 2013CB933401), the NSFC (Grant No. 11474279), and the Strategic Priority Research Program of the Chinese Academy of Sciences (Grant No. XDB07010100).

\appendix

\setcounter{equation}{0}
\setcounter{figure}{0}
%\setcounter{table}{0}
%\setcounter{page}{1}
%\makeatletter
\renewcommand{\theequation}{A\arabic{equation}}
\renewcommand{\thefigure}{A\arabic{figure}}

\section{\textit{Ab-initio} optimization principle approach in one dimension}

Understanding the 1D AOP \cite{AOP,tMPS} will be a lot of help to understand the higher-dimensional version. In this section, we present the 1D AOP scheme that does not have the Hermitian requirement \cite{tMPS}. Let us take the following translationally invariant Hamiltonian as an example, which reads
\begin{eqnarray}
\hat{H}_{Inf} = \sum_{n} \hat{H}_{n,n+1},
\label{eqs-Hinf}
\end{eqnarray}
with $\hat{H}_{n,n+1}$ the two-body interaction.

With the Hamiltonian, the next step is to prepare the cell tensor that parameterize the self-consistent equation. This step is the same for the AOP approaches introduced in Sec. III. One firstly chooses a supercell that obeys the translational invariance, e.g. two adjacent sites. The bulk interaction is simply the coupling between these two spins, i.e. $\hat{H}^B(i,j) = \hat{H}(i,j)$. Then we define the two-body operator $\hat{F}^{\partial}(i,j) = I-\tau \hat{H}(i,j)$ as the shifted interaction on the boundary of the supercell and do the singular value decomposition as
\begin{eqnarray}
	\hat{F}^{\partial}(i,j) = \sum_{a} \hat{F}^{L}(i)_a \otimes \hat{F}^{R}(j)_a.
\label{eqs-Fboundary}
\end{eqnarray}
$\hat{F}^{L}(i)_a$ and $\hat{F}^{R}(j)_a$ are two sets of one-body operators (labeled by $a$) acting on the left and right spins of $\hat{F}^{\partial}(i,j)$, respectively.

The cell tensor is defined as the product of the (shifted) bulk Hamiltonian with $\hat{F}^{L}_a$ and $\hat{F}^{R}_a$ as
\begin{eqnarray}
\hat{\mathcal{F}}(i,j)_{a_1a_2}= \hat{F}^{R}(i)_{a_1} \hat{F}^{L}(j)_{a_2} \tilde{H}^B(i,j),
\label{eqs-getF}
\end{eqnarray}
with $\tilde{H}^B(i,j) = I- \tau \hat{H}^B(i,j)$.

\begin{figure}[tbp]
	\includegraphics[angle=0,width=1\linewidth]{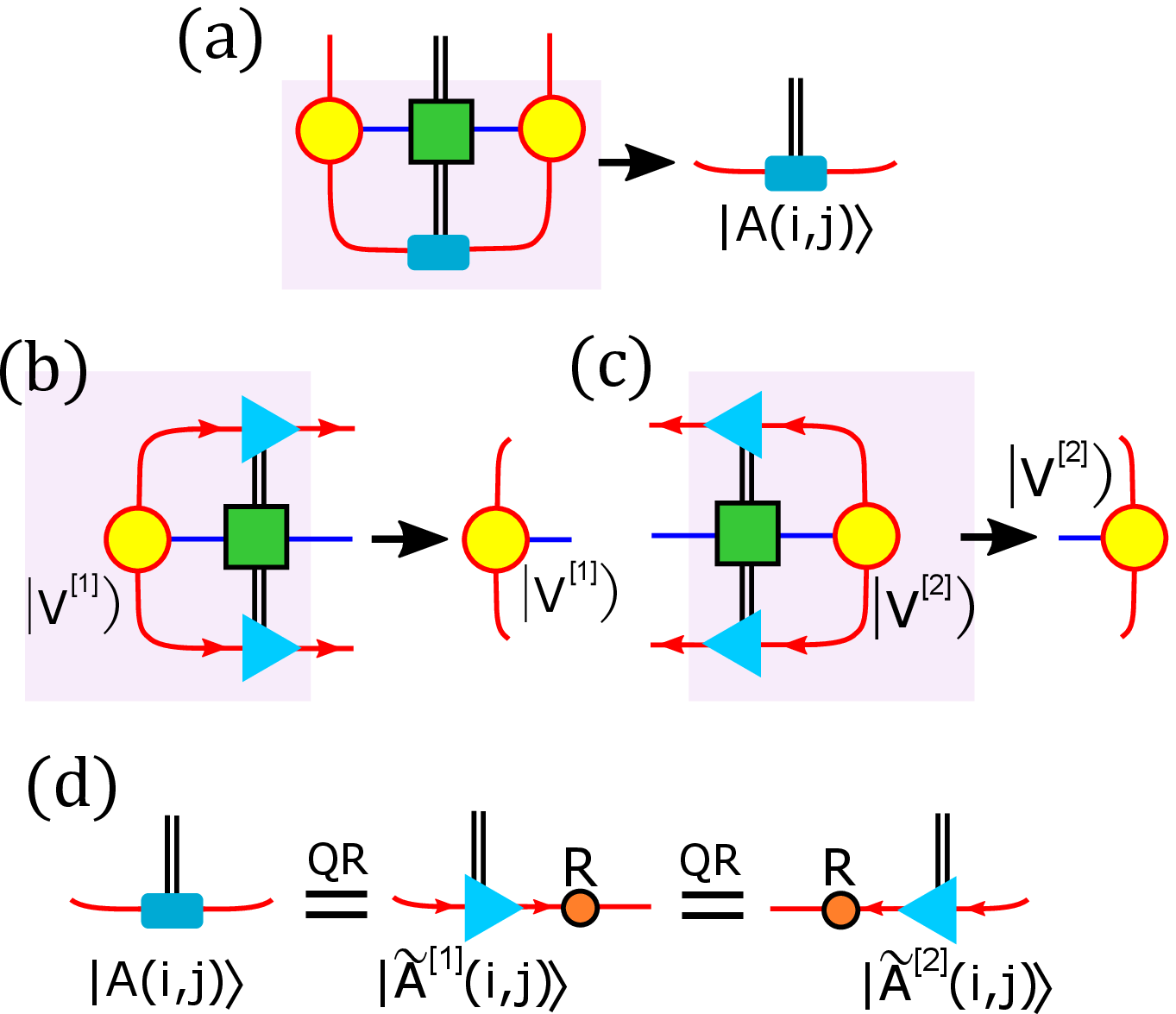}
	\caption{(Color online) The illustration of the self-consistent eigenvalue equations of 1D AOP approach.}
	\label{figs-Eigeqs}
\end{figure}

Then, with two boundary tensors $|V^{[x]})$ (guessed or previously obtained in the last iteration) and the cell tensor $\hat{\mathcal{F}}(i,j)$, we define three eigenvalue equations (Fig. \ref{figs-Eigeqs}) as
	\begin{eqnarray}
	\hat{\mathcal{H}}(i,j)_{\mu_1'\mu_2,\mu_1\mu_2} = \sum_{a_1a_2} (V^{[1]}|_{a_1\mu_1 \mu_1'} \hat{\mathcal{F}}(i,j)_{a_1a_2} |V^{[2]})_{a_2\mu_2 \mu_2'}, \label{eqs-effectH}\\
	M^{[1]}_{a_1\mu_1\mu_1',a_2\mu_2\mu_2'} = \langle \tilde{A}^{[1]}(i,j)|_{\mu_1'\mu_2'} \hat{\mathcal{F}}(i,j)_{a_1a_2} |\tilde{A}^{[1]}(i,j)\rangle_{\mu_1\mu_2},\label{eqs-M1}\\
	M^{[2]}_{a_1\mu_1\mu_1',a_2\mu_2\mu_2'} = \langle \tilde{A}^{[2]}(i,j)|_{\mu_1'\mu_2'} \hat{\mathcal{F}}(i,j)_{a_1a_2} |\tilde{A}^{[2]}(i,j)\rangle_{\mu_1\mu_2}.\label{eqs-M2}
	\end{eqnarray}

By solving the leading eigenvector of $\hat{\mathcal{H}}(i,j)$ given by Eq. (\ref{eqs-effectH}), we obtain the central tensor $|A(i,j)\rangle_{\mu_1\mu_2\mu_3\mu_4}$. The central tensor can be considered as a state in the Hilbert space of the supercell labeled by two virtual indexes. Meanwhile, $|V^{[x]})$ is obtained as the (left) leading eigenvector of $M^{[x]}$ [Eqs. (\ref{eqs-M1}) and (\ref{eqs-M2})]. $M^{[x]}$ is defined by the isometries $|\tilde{A}^{[x]}(i,j) \rangle$ that is obtained by the QR decomposition of $|A(i,j)\rangle$. For example for $x=1$, we have
\begin{eqnarray}
|A(i,j)\rangle_{\mu_1\mu_2} = \sum_{\nu} |\tilde{A}^{[1]}(i,j)\rangle_{\nu \mu_2} R^{[1]}_{\nu \mu_1}.
\label{eqs-QR}
\end{eqnarray}
$|\tilde{A}^{[1]}(i,j)\rangle$ is orthogonal, satisfying
\begin{eqnarray}
\sum_{\mu_2} \langle \tilde{A}^{[1]}(i,j)|_{\mu_1 \mu_2} \tilde{A}^{[1]}(i,j)\rangle_{\mu_1' \mu_2} = I_{\mu_1 \mu_1'}.
\label{eqs-OrtA3}
\end{eqnarray}
These isometries play the role of the renormalization group flow in the standard DMRG \cite{DMRG}. Similarly, $|V^{[x]})_{a_x\mu_x \mu_x'}$ can be understood as a ``state'' defined in the space of the boundary index $a_x$ labeled by $\mu_x$ and $\mu_x'$. One can see that these equations are parametrized by the solutions of others, and can be solved in an alternative way in practice.

\begin{figure}[tbp]
	\includegraphics[angle=0,width=1\linewidth]{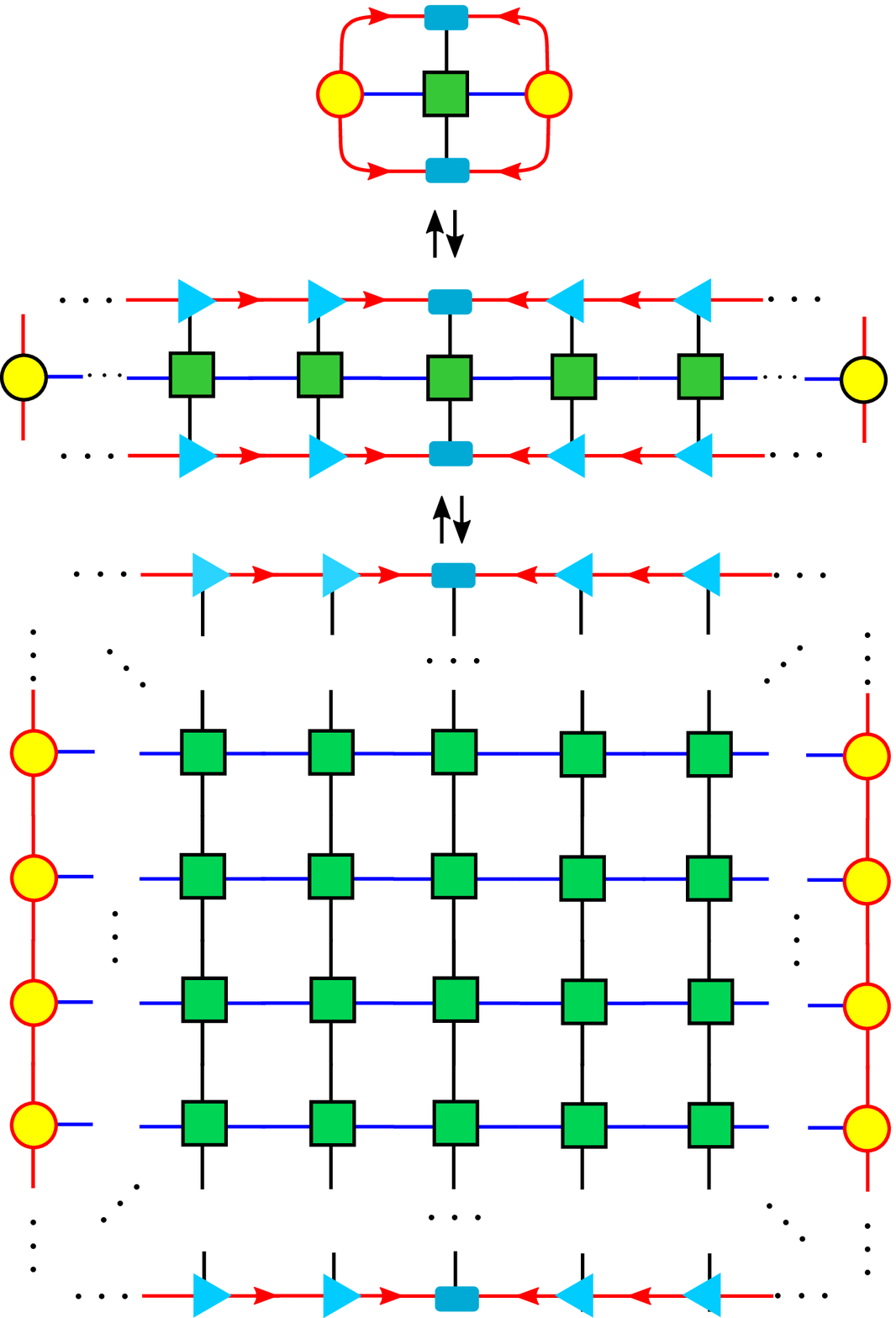}
	\caption{(Color online) The illustration of the encoding scheme for a 2D TN.}
	\label{figs-TNE}
\end{figure}

In the language of TN, Eqs. (\ref{eqs-effectH})-(\ref{eqs-M2}) encode an infinite TN that represents the imaginary time evolution for the ground state (Fig. \ref{figs-TNE}). Define a local scalar function shown in the top of Fig. \ref{figs-TNE}. It is easy to see that this function is maximized at the fixed point of the self-consistent equations. Then, one can reconstruct an infinite MPO multiplied with an MPS and its conjugate by repetitively replacing $|V^{[x]})$ by $M^{[x]}|V^{[x]})$. Still, the scalar function is maximized, meaning the MPS is the dominant eigenstate of the MPO. One can again iteratively replace one MPS by the product of the MPS and the MPO to reconstruct the whole infinite TN. Note such a reconstruction can be understood in an opposite order: by going from the bottom to the top of Fig. \ref{figs-TNE}, it actually gives a contraction scheme of the TN.

There are two important constraints to realize such a reconstruction. In the step from the local scalar function to MPO, we have $(V^{[x]}|V^{[x]})=1$ since it should be the eigenstate of $M^{[x]}$. In the second step from the MPO to the whole TN, we require that the MPS is normalized, which is actually a non-local constraint. In the original proposal of the 1D AOP \cite{AOP}, this constraint is turned to be local with some tricks under the assumption that all eigenvalue problems are Hermitian. In the generalized 1D version \cite{tMPS} presented above, the MPS is normalized because of the orthogonality of $|\tilde{A}^{[x]}(i,j)\rangle$.

Now we explain the few-body Hamiltonian that mimics the ground state of the infinite 1D chain. By reviewing Eq. (\ref{eqs-getF}), the matrix [Eq. (\ref{eqs-effectH})] whose eigenstate gives $|A(i,j)\rangle_{\mu_1\mu_2}$ can be written as the product of three parts: one bulk term and two boundary terms. The bulk term contains simply the physical interactions of the original model. Same to Eq. (\ref{eq-Hbath}) in higher-dimensional AOP, the boundary parts are defined as
\begin{eqnarray}
 \hat{\mathcal{H}}^{\partial}(n,x)_{\mu_x \mu_x'} = \sum_{a_x} \hat{F}^{L(R)}(n)_{a_x} |V^{[x]})_{a_x\mu_x\mu_x'}.
\label{eqs-Hbath}
\end{eqnarray}
Then same to Eq. (\ref{eq-Hfewbody}), the few-body Hamiltonian for the 1D simulator is obtained as
\begin{eqnarray}
\hat{H}^{FB} = \sum_{\langle i,j \rangle \in cluster} \hat{H}(i,j) + \sum_{\langle x \in cluster,n \in bath \rangle} \hat{H}^{\partial}(n,x).
\label{eqs-Hfewbody}
\end{eqnarray}
where $\hat{H}^{\partial}(n,x)$ satisfies $\hat{\mathcal{H}}^{\partial}(n,x) = I - \tau \hat{H}^{\partial}(n,x) + O(\tau^2)$, same to Eq. (\ref{eq-Hexp}).

Comparing the 1D and higher-dimensional AOP versions, we can see many connections, including the idea of defining the eigenvalue equations, the constraints for the encoding/reconstruction process, and the emergence of the few-body Hamiltonians. The differences are also crucial. For 1D quantum systems, we can directly encode the imaginary-time-evolution TN. In higher dimensions, the tree approximation is introduced. The essential reason is to satisfy the second constraint, which is the normalization of the state ansatz (see Appendix B). The normalization of a standard TN state (e.g., on square lattice) requires an extra loop of TN encoding or contraction. The AOP algorithm to directly encode the ($D+1$)-dimensional imaginary-time-evolution TN for $D$-dimensional quantum system ($D\geq 2$) is still an open issue. This will lead to a general form of the few-body Hamiltonians (see Appendix E).

\section{State ansatz behind our approach in higher dimensions}

At the first stage, the ansatz is an infinite tree PEPS that optimally approximates the ground state in the \textit{rank-1} sense \cite{Rank1,NCD}. This can be seen from the tensor network (TN) encoded in the self-consistent eigenvalue equations. Starting from Eq. (\ref{eq-effectH}), one can substitute each of the boundary tensors $|V^{[x]})$'s by the contraction of the other three $|V^{[x]})$'s, $|\tilde{A}^{[x]})$, $(\tilde{A}^{[x]}|$ and the cell tensor $\hat{\mathcal{F}}$ according to Eqs. (\ref{eq-M1})-(\ref{eq-M4}). We are using the fact that $|V^{[x]})$ is the eigenvector of $M^{[x]}$. By doing so repetitively, an infinite tree PEPS formed by $|A\rangle$ and $|\tilde{A}^{[x]})$'s can be grown to reach the thermodynamic limit. At the same time, the TN that gives ($I-\tau \hat{H}_{tree}$) appears, where $\hat{H}_{tree}$ is the Hamiltonian defined on the tree. The local interactions of $\hat{H}_{tree}$ are exactly the same with the original model as long as one only looks at a loop-free subsystem, thus $\hat{H}_{tree}$ provides a reasonable approximation. Such a tree PEPS minimizes the energy of $\hat{H}_{tree}$.

For better understanding the approximation of the state on, e.g., an infinite square lattice, we could ``grow'' the tree in such a way that it fills the whole square lattice. Inevitably, some $|V^{[x]})$'s on the boundary of the tree will gather at the same site. The tensor product of these $|V^{[x]})$'s in fact gives the optimal rank-1 approximation \cite{Rank1} of the tensor that forms the bulk of tree TN (translational invariant). Now, if one uses the full-rank tensor to replace its rank-1 version (the tensor product of four $|V^{[x]})$'s), one will have the TN of $I- \tau \hat{H}$ (with $H$ the target Hamiltonian on square lattice) instead of $I-\tau \hat{H}_{tree}$, and the tree PEPS becomes the one defined on the square lattice. Such a picture can be understood in the opposite manner: imaging that one has the ``correct'' TN defined on the square lattice, what we do is to replaced certain tensors by its rank-1 approximations to destruct all the loops of the TN. In this sense, the tree PEPS defined on the original lattice (not actually a Cayley tree or Bethe lattice \cite{CayleyTree,Bethe}) in stage one provides the optimal loop-free approximation of the ground state, where the loops are destructed by the rank-1 tensors. It would be very helpful to refer to the figures and the discussions in Ref. [\cite{NCD}] that are given considering TN contractions.

There are several issues we shall stress. Firstly, one will actually \textit{not} do the above substitutions to reconstruct the tree PEPS. It is automatically encoded in the self-consistent equations. The ``reconstruction picture'' is proposed only to understand the ansatz behind the approach. Secondly, one may notice that the self-consistent equations proposed here are slightly different from those for the rank-1 decomposition of a single tensor \cite{Rank1}. The reason is that in our case, the normalization of the PEPS should be considered when doing the rank-1 approximation. We here borrow the idea of iDMRG on the tree PEPS \cite{TTN1,TTN2} to satisfy this constraint. The third issue is about the uniqueness of the reconstruction of the tree PEPS. Indeed, the contraction of three $|V^{[x]})$'s, $|\tilde{A}^{[x]})$, $(\tilde{A}^{[x]}|$ and $\hat{\mathcal{F}}$ to substitute $|V^{[x]})$ is not unique. However, it is unique when we require the presence of $|\tilde{A}^{[x]})$, $(\tilde{A}^{[x]}|$ and $\hat{\mathcal{F}}$, in order to recover the TN's of $I-\tau \hat{H}$ as well as the tree PEPS. This is due to the uniqueness of the rank-1 decomposition, which is argued to be a concave problem \cite{Rank1}.

Such a tree approximation is also closely related to the iPEPS algorithms called simple update \cite{SimpleUpdate,ODTNS,SimpleTN3D}, where the infinite PEPS is updated by considering the local environment. After reaching the fixed point, the PEPS satisfies a set of self-consistent equations, which lead to a similar tree structure \cite{ODTNS}. Even some long-rang effects are ignored, simple update are still quite accurate especially for gapped states.

Aimed at reducing the error of the tree approximation, the second stage of our approach is to construct the few-body Hamiltonian $\mathcal{\hat{H}}^{FB}$ on a larger cluster by reusing the bath obtained in the first stage, and then calculate the ground state of $\mathcal{\hat{H}}^{FB}$ with a finite-size algorithm. The ansatz behind can be considered as a generalized tree PEPS. In the center of the PEPS, the tensor contains all the physical sites inside the cluster, connected with several infinite tree brunches that are the same to those appearing in stage one. The bath sites carry the entanglement between the physical sites in the cluster and these infinite tree brunches.

\section{``Finite-loop'' effects}
Thanks to the infinite tree brunches in the PEPS ansatz, our algorithm does not suffer the conventional finite-size effect in the algorithms such as ED, QMC or DMRG. Thus, the effects from the finiteness of the cluster in the second stage are essentially different. In the first stage, the system size is already infinite because the bath encodes the information of an infinite tree in the eigenvalue equations. Only the loops beyond the supercell are destroyed in an optimal manner (rank-1 approximation of the tensors) \cite{NCD}. In stage two, there will be no tree error inside the cluster since all interactions there are fully considered. If the cluster contains larger loops than the cell tensor used in stage one, the precision will be improved. On the other hand, there will be no improvement if one increases the size of the cluster without having larger loops. For this reason, the ``finite-size effects'' of AOP mean the errors caused by the finiteness of the considered loops.

\section{Computational cost}
The motivation to use the tree approximation is its efficiency especially for 3D quantum models. The computational cost of the first stage is that of the generalized DMRG on an infinite tree PEPS \cite{TTN1,TTN2}, which roughly scales as $\mathcal{O}(d^{2N_0} D^{3z})$ with $d$ the dimension of the physical Hilbert space on one site, $N_0$ the number of physical sites in the supercell, $D$ the dimension of a virtual index and $z$ the coordination number of the lattice \cite{NoteCost}.

To solve the few-body Hamiltonian, the computational cost (leading term) with ED scales as $\mathcal{O}(d^N D^{N^{\partial}})$ ($N$ and $N^{\partial}$ the number of physical and bath sites, respectively), and that with DMRG scales as $\mathcal{O}[(N+N^{\partial})\max(d,D)^3\chi^6]$ ($\chi$ the bond dimension cut-off of DMRG). The cost is similar to solving a nearest-neighbor finite-size system that contains two kinds of sites, whose local Hilbert space is of dimension $d$ (physical) and $D$ (bath), respectively. Surely one can choose other algorithms to solve the few-body Hamiltonian in the second stage, such as QMC or finite PEPS algorithms \cite{FinitePEPS11,FinitePEPS12}. Benefits from the fact that the few-body Hamiltonian is the product (or summation) of local couplings, the efficiency will be similar to that when applying to the standard (short-range) Hamiltonians. In addition, it is possible to update the bath simultaneously in stage two, and the computational cost would be approximately identical to the cluster update schemes of TN.

\begin{figure}[tbp]
	\includegraphics[angle=0,width=1\linewidth]{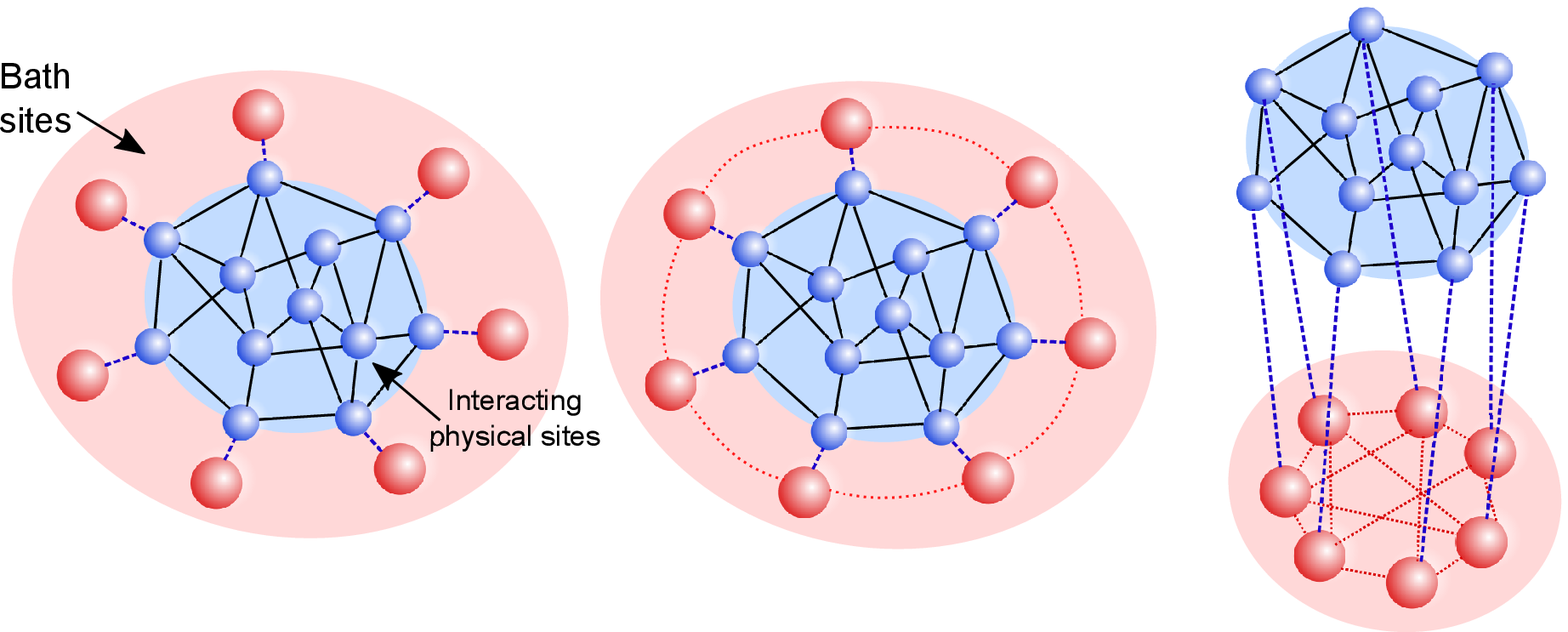}
	\caption{(Color online) The illustrations of three kinds of possible few-body Hamiltonians that contain several interacting physical and bath sites. All the physical interactions (black lines) inside the chosen cluster should be fully considered. The left figure illustrates the one by using the tree DMRG for the physical-bath interactions $\hat{\mathcal{H}}^{\partial}$ (blue dashes), where there are no bath-bath interactions. By choosing other algorithms (e.g., SRG or CTMRG) to calculate $\hat{\mathcal{H}}^{\partial}$, it is possible to also have nearest-neighbor (middle figure) or even long-range (right figure) bath-bath interactions (red dots).}
	\label{fig-Baths}
\end{figure}

\section{General forms of few-body Hamiltonian}
As discussed above, the dominant error comes from the destruction of the loops. As a consequence, the interactions between the bath and the physical sites are the tensor product of local terms
\begin{eqnarray}
\hat{\mathcal{H}}^{FB} = \prod_{\langle x,n \rangle} \hat{\mathcal{H}}^{\partial}(n,x).
\end{eqnarray}
It means that in the standard summation form, there are no bath-bath interactions (Fig. \ref{fig-Baths}). The tree branches in the ground-state ansatz are not connected to each other from anywhere else than the central part.

One can adopt other TN algorithms such as the cluster or full update schemes \cite{PEPS11,PEPS12,PEPScluster11,PEPScluster12,CTMRG0,CTMRG11,CTMRG12,VarPEPS,GrdPEPS,SRG,HOSRG} to obtain the physical-bath interactions. Then the Hamiltonian will not simply be the tensor product, but generally given by
\begin{eqnarray}
\hat{\mathcal{H}}^{FB} = \sum_{\{\alpha\}} \prod_{\langle x,n \rangle} \hat{\mathcal{H}}^{\partial}(n,x)^{\alpha_{x,n}}.
\end{eqnarray}
Then the bath-bath interactions will appear in the standard summation form. See the illustrations of three possible situations in Fig. \ref{fig-Baths}. The extra summations will lead to another (similar) PEPS ansatz beyond the one with tree branches, which should better mimics the infinite environment. However, the computational cost with the currently known methods will become much more sensitive to the coordination number and the dimensionality of the model, making the 3D ground states extremely difficult to access.

\section{Discussions about imaginary-time evolution picture and criticality in higher dimensions}

The idea of approximating an infinite Hamiltonian with a finite effective one has been proposed for the time evolution of 1D quantum systems \cite{IBMPS}. An important difference in our work is that the ``evolution'' of the finite effective model is constructed not from a new $\hat{H}$ but with a shift ($I-\tau \hat{H}$) that is in fact the imaginary-time evolution operator. It brings several operational advantages for simulating the ground states, in particular, of higher-dimensional systems. The triangular structure of the Hamiltonian is avoided here, thus the eigenvalue equations for the boundary tensors have stable solutions and the entanglement bath is well-defined. The few-body Hamiltonians with the bath of higher-dimensional systems can be easily constructed as the summations of local terms.

Our scheme makes it possible to adopt the ($1+1$)-D scaling theories for characterizing criticality \cite{EntCritic} to higher-dimensional models. It is known that any TN algorithms, essentially, cannot give directly a divergent correlation length at the critical point. For 1D quantum systems, it has been shown that at the critical point, any MPS with a finite bond dimension is gapped and possesses a finite correlation length $\xi$ \cite{EntCritic} satisfying
\begin{eqnarray}
\xi \sim \mathcal{D}^{\kappa},
\end{eqnarray}
with $\mathcal{D}$ the bond dimension of the MPS and $\kappa$ its scaling exponent. One can see that with a finite $\mathcal{D}$, $\xi$ is always finite, and the information of the criticality is in hidden the algebraical scaling behavior when $\mathcal{D}$ increases. For the scaling of magnetic field $h$ near the critical point, the algebraic behavior of $\xi$ versus $h$ can still survive, however, the value of the exponent might be inaccurate.

For a 2D PEPS, one has to compute the contraction of a 2D TN (e.g., by iTEBD with MPS) to get its correlations using finite dimension cut-offs, and thus the results will still be finite \cite{EntCritic2D}. To tackle this difficulty, it has been proposed that the divergence of the correlations can be studied by the scaling of the bond dimension of the MPS, from which the central charge of the conformal field theory to characterize the criticality can be accurately obtained \cite{EntCritic,EntCritic2D}.

In our approach, the dynamic correlation length of the ground state $\xi$ is given by the correlation length of an infinite MPS formed by $|V^{[x]})$ in the imaginary time direction, written as $|\tilde{\psi} \rangle = \sum_{\{\mu\}} \prod |V^{[x]})_{a_x\mu_x\mu_x'}$. Such an MPS (dubbed as time MPS) is quasi-continuous (discretized up to the Trotter step $\tau \to 0$). Let us explain how to get $\xi$ in the AOP approach. In higher dimensions, the scheme is similar.

\begin{figure}[tbp]
	\includegraphics[angle=0,width=1\linewidth]{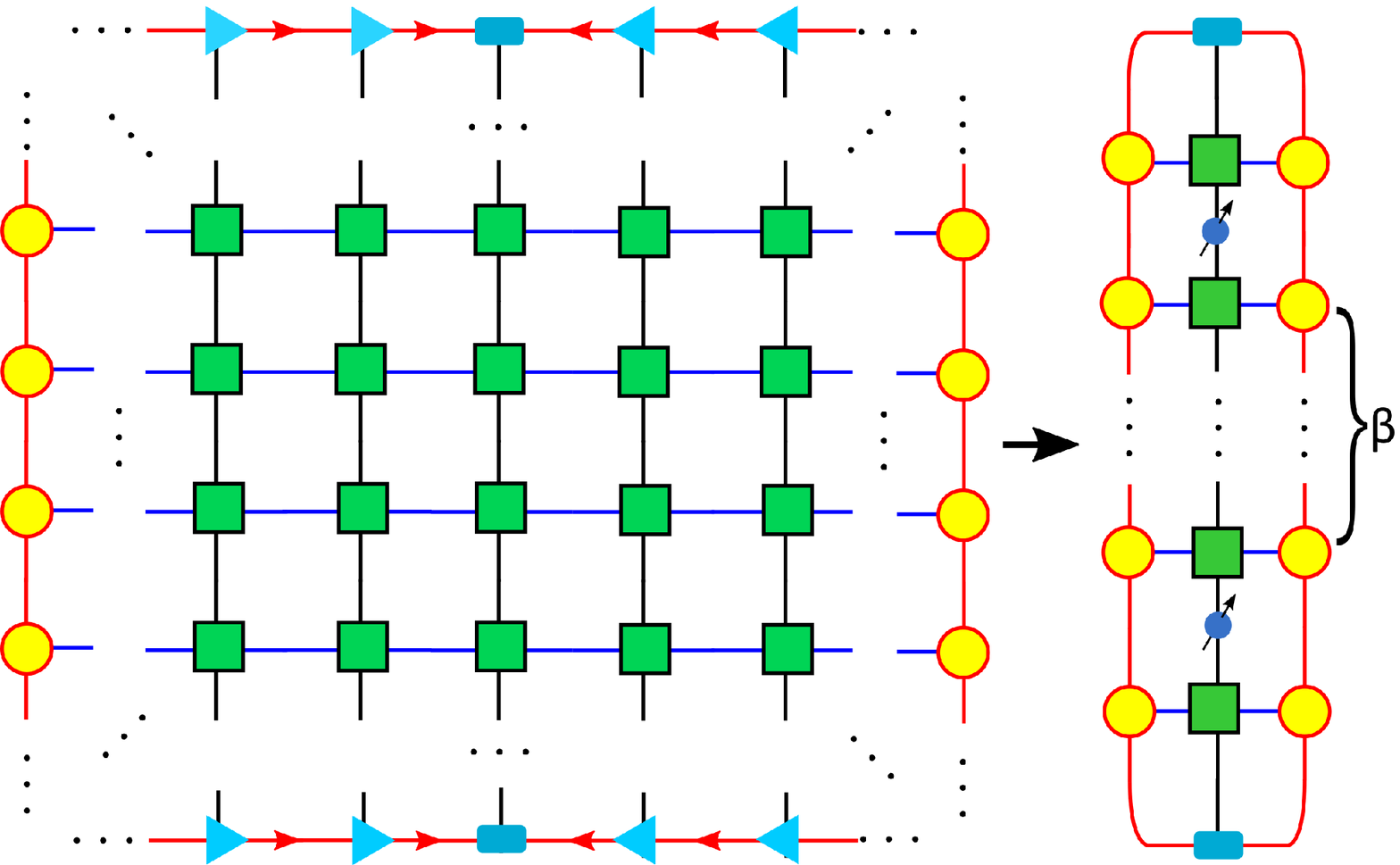}
	\caption{(Color online) The illustrations of the computation of the correlation functions in 1D AOP.}
	\label{figs-corr}
\end{figure}

The dynamic correlation function of the ground state is defined as $\langle \Phi |\hat{S} e^{-\beta \hat{H}} \hat{S}|\Phi \rangle/e^{-\beta E} - \langle \Phi|\hat{S}|\Phi \rangle^2$ with $|\Phi \rangle$ the ground state and $E$ the ground-state energy. In our framework, it is the contraction of a TN, where the two operators are put in the same column. Thanks to the encoding scheme, such a contraction becomes the contraction of a tensor stripe (Fig. \ref{figs-corr}). This stripe is the product of $\mathcal{\hat{H}}^{FB}$'s [Eq. (\ref{eq-Hfinite}), also called the transfer matrix] with the two operators in between. The dynamic correlation length is defined as the exponent of the correlation length. One can see that such an exponent is obtained simply by
\begin{eqnarray}
\xi=\frac{\tau}{\log \Lambda_0 - \log \Lambda_1},
\end{eqnarray}
with $\Lambda_0$ and $\Lambda_1$ the two largest eigenvalues of $\mathcal{\hat{H}}^{FB}$. 

An advantage of the dynamic correlation properties is that we find much less finite-loop or finite-dimension-cutoff effects than the spatial correlations. This is also supported by a recent DMRG work \cite{DMRGscaling}, where the finite-size effects are found to be much smaller for the dynamic correlations. Meanwhile, a finite dimensional matrix cannot give a critical spectrum. It means one cannot directly obtain a divergent correlation length at the critical point, and a scaling of the dimension would be necessary to identify the criticality. How to do such kind of scalings for 2D and 3D states is still an open question.

\section{Relations to other algorithms}
\begin{figure}[tbp]
	\includegraphics[angle=0,width=1\linewidth]{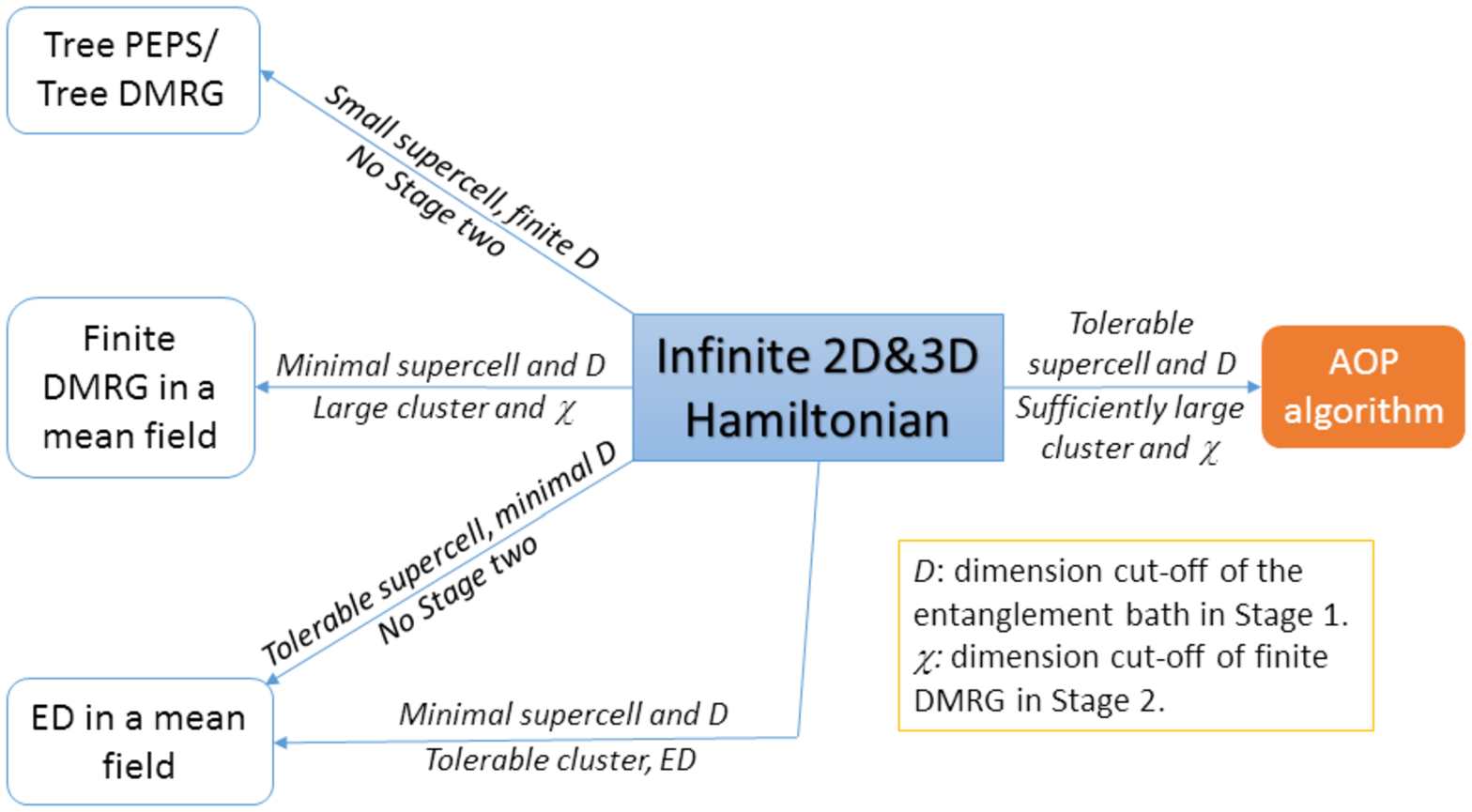}
	\caption{(Color online) Relations between AOP and several existing algorithms (PEPS, DMRG and ED) for the ground-state simulations of 2D and 3D Hamiltonian. The corresponding computational set-ups in the first (bath calculation) and second (solving the few-body Hamiltonian) stages of AOP algorithm are given above and under the arrows, respectively.}
	\label{fig-AOPrelation}
\end{figure}

By taking certain limits of the computational parameters, the relations among our approach and other algorithms are illustrated in Fig. \ref{fig-AOPrelation}. The simplest situation is to take the dimension of the bath sites $\rm dim (\mu_x)= \rm dim (\mu_x') =1$, and then $\hat{\mathcal{H}}^{\partial}$ can be written as a linear combination of spin operators (and identity). Thus in this case, $|V^{[x]})$ simply plays the role of a mean field. If one only uses the bath calculation of the first stage to obtain the ground-state properties, the algorithm will be reduced to the tree DMRG \cite{TTN1,TTN2}. If one takes the minimal supercell with $D=1$ in stage one, the entanglement bath will be reduced to a magnetic mean field. By choosing a large cluster, the DMRG \cite{DMRG} simulation in stage two becomes equivalent to the standard DMRG for solving the cluster in a mean field. If one uses $D=1$ and chooses a supercell of a tolerably large size in the first stage without entering stage two, or if one chooses a small cluster with $D=1$ in stage one and uses ED in stage two to solve the few-body Hamiltonian with a tolerably large cluster, our approach will become the ED on the corresponding finite system in a mean field. By taking proper supercell, cluster, algorithms and computational parameters, our approach outperforms others.

\section{Generalization to ($d \geq 4$) dimensions}
Benefiting from its flexibility, it is possible to generalize our approach to even $(d \geq 4)$-dimensional quantum models. The main problem to be tackled is the computational cost. In the second stage by using DMRG for example, the cost increases polynomially with the size of the cluster, thus also polynomially with the dimensionality $d$. In the first stage with tree DMRG, the cost increase exponentially with $d$, which makes the simulations for higher-dimensional models extremely expensive. Luckily, the main task here is to solve ($2d+1$) number of self-consistent eigenvalue equations, say five [Eq. (\ref{eq-effectH})-(\ref{eq-M4})] for 2D, seven for 3D and nine for 4D quantum systems. One way to lower the cost from exponential to polynomial expenses is to use a finite algorithm such as DMRG to solve each eigenvalue problem. It is certain that the stability and efficiency have to be tested.

\section{Open issues}
Several following-up issues are to be further investigated. The flexibility allows for possible incorporating with other methods. For example, the TN techniques with symmetries \cite{DMRGsymme11,DMRGsymme12,Qspace} can be introduced to lower the computational cost so that much larger clusters can be reached in the second stage. Besides the tree DMRG \cite{TTN1,TTN2}, the other TN optimization schemes such as TN variational techniques \cite{Fupdate,GrdPEPS,VarPEPS} and tensor renormalization group algorithms \cite{TRG,CTMRG11,CTMRG12,SRG,HOSRG,LoopTN} can be adapted when the cost is tolerable. The finite-size scaling of the cluster should be explored. Our approach could also be readily generalized to higher-dimensional bosonic and fermionic lattice models. The entanglement embedding idea with the physical-bath Hamiltonian proposed here can be adopted to develop novel algorithms for infinite systems by hybridizing with other methods such as QMC, finite or tree PEPS algorithms \cite{TTNcluster,FinitePEPS11,FinitePEPS12}, or the approaches in material sciences and quantum chemistry, such as DFT \cite{DFTrev} and DMET \cite{DMET}.

%\input{AOP3Dscifile_MainPlusSM.bbl}
%\bibliographystyle{apsrev4-1}
%\bibliography{AOP3Dbib}
%\bibliographystyle{unsrt}
%\bibliographystyle{Science}

%\bibliography{AOP3Dbib}Das Sarma

\begin{thebibliography}{99}
\bibitem[{Hua()}]{Huang}
	\bibinfo{note}{We are indebted to Yi-Zhen Huang for providing these QMC data,
		where the worm algorithm is used with the inverse temperature around 50}.
\bibitem[{Not({\natexlab{b}})}]{NoteIndex}
	\bibinfo{note}{In the following, we simplify the indexes when mentioning a
		tensor if it causes no confusion.}
\bibitem[{Not({\natexlab{c}})}]{NoteBracket}
	\bibinfo{note}{To distinguish from the states in physical Hilbert space
		$|\ast\rangle$, the symbol $|\ast)$ is used to denote a state in the space of
		boundary indexes $\{a\}$.}
\bibitem[{Not({\natexlab{d}})}]{NoteCost}
	\bibinfo{note}{This is actually the leading term of the cost without using any
		tricks, such as taking advantage of the sparse property.}
\end{thebibliography}
%\bibliographystyle{unsrt}

\end{document}